# THE ANCIENT EGYPTIAN COSMOLOGICAL VIGNETTE: FIRST VISUAL EVIDENCE OF THE MILKY WAY AND TRENDS IN COFFIN DEPICTIONS OF THE SKY GODDESS NUT

**Or Graur**

*Institute of Cosmology and Gravitation, University of Portsmouth, Portsmouth PO1 3FX, UK, and Department of Astrophysics, American Museum of Natural History, Central Park West and 79th Street, New York, NY 10024-5192, USA.*

E-mail: or.graur@port.ac.uk

**Abstract:** Here, I test the long-held assumption that the ancient Egyptian Sky Goddess Nut represented the Milky Way by examining Nut's visual depictions on ancient Egyptian coffins. I assemble a catalog of 555 coffin elements, including 118 cosmological vignettes from the 21st/22nd Dynasties, and report several observations. First, the cosmological vignette on the outer coffin of Nesitaudjatakhet bears a unique feature: a thick, undulating black curve that bisects Nut's star-studded body and recalls the Great Rift that cleaves the Milky Way in two. Similar undulating curves bisect the astronomical ceiling in the tomb of Seti I and appear as part of depictions of Nut in the tombs of Ramesses IV, VI, and IX. Moreover, the undulating curve resembles similar features identified as the Milky Way on the bodies of Navajo, Hopi, and Zuni spiritual beings. Hence, I argue that the undulating curve is a visual representation of the Milky Way and that it supports a previously suggested identification *of mr-nḫ3* (Winding Waterway) as the Galaxy's Egyptian name. However, the rarity of this feature strengthens the conclusion reached by Graur (2024a): Though Nut and the Milky Way are linked, they are not synonymous. Instead of acting as a representation of Nut, the Milky Way is one more celestial phenomenon that, like the Sun and the stars, is associated with Nut in her role as the sky. Second, Nut's body is decorated with stars in only a quarter of the cosmological vignettes, suggesting that the Egyptians of the 21st/22nd Dynasties may have had a marked preference for the day sky over the night sky. Finally, I discuss the interplay between Nut's cosmological vignette and full-length portraits inside coffins from the New Kingdom to the Roman Period in the context of Nut's combined cosmological and eschatological roles as an embodiment of the coffin.

## 1 INTRODUCTION

The astronomy and cosmogony of ancient Egypt have been active subjects of research since the inception of Egyptology (for a recent review, see Belmonte and Lull, 2023). Several independent cosmogonies developed in different parts of Egypt: Heliopolis, Memphis, Thebes, Hermopolis, and Esna (for a detailed discussion, see Belmonte and Lull, 2023: 1–22). In the Heliopolitan model, the world is surrounded by an abyss of endless, inert water. Out of this abyss, the god Atum created himself. He then created Shu and Tefnut, the God of the atmosphere and the Goddess of moisture. Both siblings and lovers, Shu and Tefnut gave birth to Geb and Nut, Earth and Sky. In turn, Geb and Nut gave birth to the brothers Osiris and Seth and the sisters Isis and Nephthys. Together, these nine gods are referred to as the Ennead. Though different cosmogonies and deities came and went out of fashion throughout Egyptian history, the gods of the Ennead remained important figures in Egyptian religion and cosmology.

Various Egyptian gods are either associated with, symbolize, or directly embody certain celestial objects. The Sun, most famously, is personified as several major gods: the scarab-headed Khepri is thought to personify the rising Sun; the falcon-headed Re is the midday Sun; and Atum, head of the Ennead, is the setting Sun (Allen, 1988: 10). Similarly, the Moon is associated with the God Khonsu, while Sirius is associated with the Goddess Sopdet (also known as Sothis), and certain stars within the constellation Orion are linked to the God Sah.

Several attempts have been made to link Nut (*nwt*) with the Milky Way (Davis, 1985; Graur, 2024a; Kozloff, 1992; 1994; Wells, 1992; for a discussion of these works, see Graur, 2024a). In the latest such attempt, Graur (*ibid.*) showed that the textual description of Nut's arms in the *Fundamentals of the Course of the Stars* (the original name of the book known popularly as the '*Book of Nut*'; for an in-depth translation and discussion of the *Fundamentals* … see von Lieven, 2007) was consistent with the Milky Way's appearance in the Egyptian night sky during the winter months. Graur (2024a) also noted that Nut's roles in the transition of the deceased to the afterlife (as described in spells from the *Pyramid Texts* and *Coffin Texts*) and her link to the annual bird migration (as described in the *Fundamentals* …) were similar to roles attributed to the Milky Way by other cultures (for comparative studies of the Milky Way among different cultures see, e.g., Berezkin, 2010; Graur, 2024b; Krupp, 1991; 1995; Lebeuf, 1996). The main point made by Graur (2024a), as opposed to previous works, was that while Nut and the Milky Way may have been linked, they were not synonymous with each other. Instead, Graur (*ibid.*) suggested that the Milky Way served as a figurative marker of Nut's existence as the sky, highlighting her arms during the winter months and her torso or backbone during the summer. In this paper, I study Nut's cosmological vignette as it appears on ancient Egyptian coffins and identify the first visual depiction of the Milky Way; a similar effort to study Nut's depictions in funerary papyri is ongoing and will be the subject of a future paper.

## 1.1 Depictions of Nut on Ancient Egyptian Coffins

Nut's depictions on Egyptian coffins stem from the myriad roles played by this Goddess in the transition of the deceased to the afterlife (for a review, see Hollis, 2019). Several spells from the *Pyramid Texts* (carved on the walls of the burial chambers in the pyramids of the Old Kingdom) and the *Coffin Texts* (found on Middle Kingdom coffins) call on Nut to shield the deceased with her arms, reassemble their bones, nourish them, give birth to them once more, and guide them up to the sky. Graur (2024a) argued that the latter role, which is expressed as Nut reaching her arm out to the deceased or acting as a ladder, helps to link Nut to the Milky Way. Here, I focus on Nut's protection of the deceased and her identification with the coffin. These roles are attested at least as early as the Old Kingdom in several spells from the *Pyramid Texts* (see Graur, *ibid.*, and Hollis, 2019 for examples). Perhaps the most relevant of these is spell 364. Faulkner (1969: 119) translated the relevant part of the spell as:

> … you having been given to your mother Nut in her name of 'Sarcophagus', she has embraced you in her name of 'Coffin', and you have been brought to her in her name of 'Tomb'.

In his more modern translation, Allen (2015: 84) translated the same section as:

> You have been given to your mother Nut in her identity of the entombment, she has collected you in her identity of the tomb chamber, and you have been elevated to her in her identity of the tomb's superstructure.

The identification of Nut with the coffin, sarcophagus, and tomb is seen in her frequent and varied depictions on coffins from the New Kingdom down to the Roman Period. Her four most frequent visual depictions are:

(1) as a winged, clothed woman depicted in a kneeling position with her wings spread out in protection of the deceased;
(2) in a full-length, anthropomorphic portrait that takes up most of the space on the floorboard of the coffin or the interior of its lid;
(3) as a clothed woman perched in a sycamore tree, offering food and drink to the deceased and their *ba* (part of the Egyptian conception of the self, often depicted as a human-headed bird); and
(4) as part of the cosmological vignette.

In this work, I focus primarily on Nut's fourth depiction, but I devote a short discussion to her winged and full-body depictions in Section 5. For an in-depth discussion of her role as one of the Goddesses of the Sycamore, see Billing (2002).

## 1.2 Nut's Celestial Roles and the Cosmological Vignette

As the Goddess of the Sky, Nut played several roles in ancient Egyptian astronomy and cosmology. First, as the sky she protected the Earth from being inundated by the encroaching waters of the void. Second, as described in the *Fundamentals …*, Nut was an integral part of the daily Solar cycle. As the Sun set in the west, it was swallowed by Nut. It then traveled through Nut's body, which, at least for a time, was synonymous with the netherworld (Duat; *dw3t*) where it was rejuvenated in an encounter with Osiris. Nut then gave birth to the Sun as it rose in the east. In a variation of this role, Nut also gave birth to and swallowed the decan stars (certain stars and asterisms thought to have been used as a star clock during the night; see Belmonte and Lull, 2023: 123–132) as they rose and set throughout the night (von Lieven, 2007; 2010).

Nut's celestial roles gained a visual depiction in the cosmological vignette, in which she is shown as a naked, arched woman, sometimes covered with stars (Figure 1a–b). She is usually held aloft by her father Shu, while her brother Geb lies underneath. Nut's arched posture is seen as evoking her identification with the sky and its protection of the Earth below (e.g., Niwiński, 2011; te Velde, 1977). Nut's role in the daily Solar cycle is also prominent in these vignettes. She is either shown swallowing and giving birth to the Sun or carrying the Solar Bark on her back (e.g., Figure 1c–e). Geb is usually colored green or shown covered by green reeds, strengthening his role as the Earth. In peculiar cases (e.g., Figure 1f), Shu is replaced either by a headless mummy or by a different deity, e.g., the ram-headed

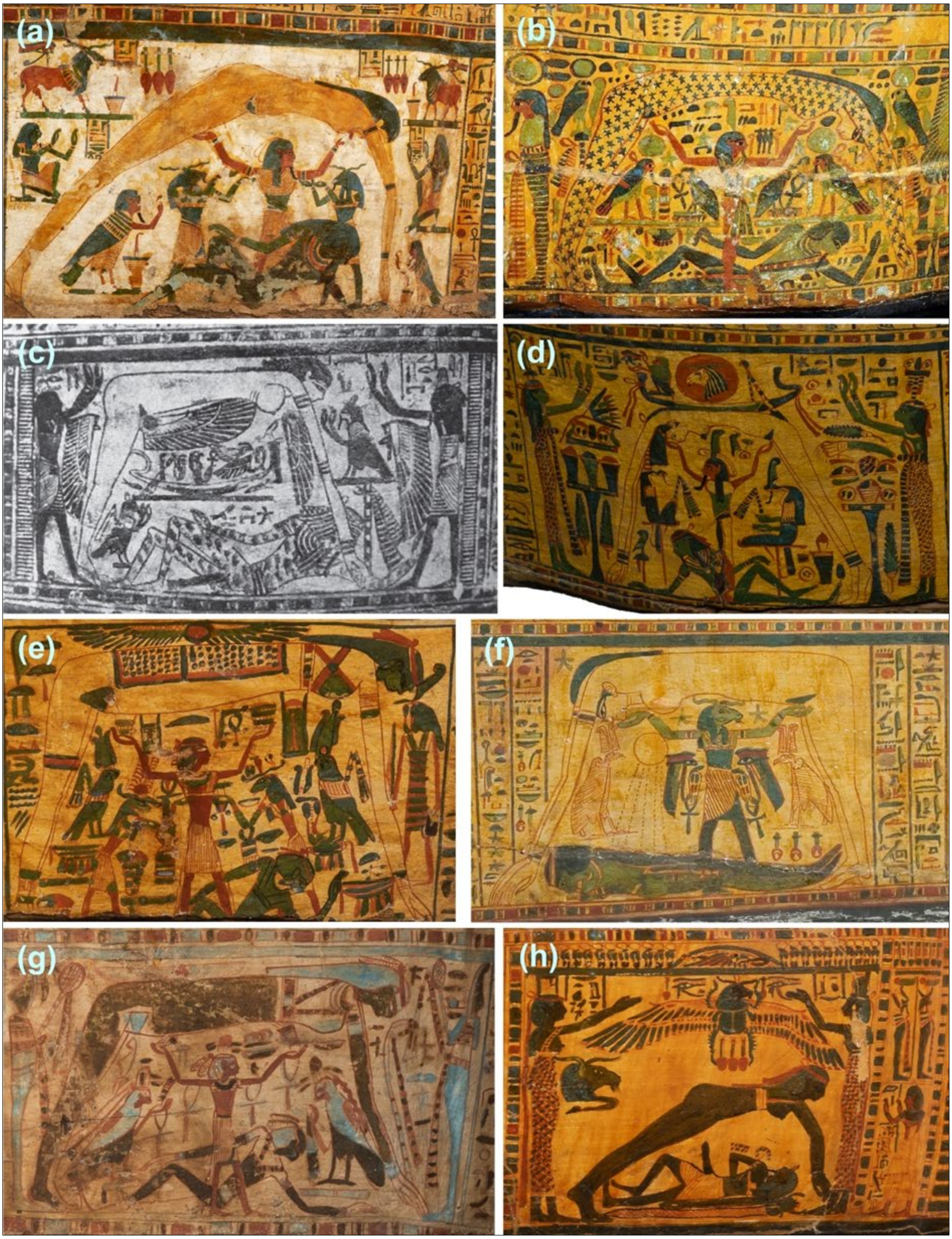

Figure 1: Examples of Nut's cosmological vignette. (a): C78, a typical example featuring a naked Nut supported by Shu and arched over a green-colored Geb (courtesy: Museo Egizio); (b) C84, a typical example in which Nut is covered in stars (courtesy: Rijksmuseum van Oudheden); (c): C89, Nut swallowing and giving birth to winged Sun discs (courtesy: Turaev, 1914); (d) C94, Nut carrying the Sun in the Solar Bark (courtesy: Museum Gustavianum); (e) C104, Nut wearing a spotted rug (courtesy: C77, Museo Egizio); (f) Nut supported by Re instead of Shu (courtesy: image © National Museums Scotland); (g) C75, an example of a black-colored Nut (courtesy: Museo Egizio, © Photo Nicola Dell'Acquila/Museo Egizio); (h) C79, example of a black-colored Nut and an ithyphallic Geb (courtesy: Mueso Egizio).

Re, the God of Magic Heka, or the dwarf god Bes. Nut is usually not colored in, except for rare instances where she is either colored white or completely black (Figure 1g–h). In one peculiar instance, Nut's head is replaced with that of a serpent (Niwiński, 1995: Figure 1).

The cosmological vignette is first attested at the beginning of the 19th Dynasty (the early Ramesside Period) in the monumental witness of the *Fundamentals* … in the Cenotaph of Seti I at Abydos (Frankfort, 1933: Plate 81). Additional witnesses of the *Fundamentals* …, which also include this vignette, are found in the tombs of Ramesses IV and Mutirdis (in all three of which Geb is absent), and the Osirian Chapel at the Temple of Dendera from the Ptolemaic Period (Cauville, 1997: Plates 204, 235; in which Shu is absent). Similar depictions of Nut as an arched woman, yet without the presence of both Shu and Geb, appear in monumental versions of the *Book of the Day* and the *Book of the Night* in the tombs of Ramesses VI and Ramesses IX (Maravelia, 2003), on funerary stele from the 22nd to the 26th Dynasties (Calmettes, 2024: 113–115), and on the walls of the Temple of Hathor and the Osirian Chapel in Dendera (Cauville, 1997: plates 60, 86, 115, 144, 260, 283). The *Book of the Divine Cow*, attested in the tombs of Seti I, Rameses III, Rameses VI, and Tutankhamun, provides a variation of the cosmological vignette in which Nut is transformed into a cow, held aloft by Shu and several Heh Gods, in order to carry the weary Re on her back (Clagett, 1992: 539–540; Hornung, 1982: 41–42; Simpson, 2005: 290–293).

However, the most prolific use of the cosmological vignette is found on coffins and funerary papyri from the 21st and early 22nd Dynasties (ca. 1075–945 and 945–715 BCE, respectively; see Table 1 for the dates of all other dynasties mentioned in this work). The burial assemblages of the 21st and early 22nd Dynasties were composed of nested anthropoid coffins: a large outer case with a mummiform lid, a smaller inner case with its own mummiform lid, and a mummy board that rested on top of the mummy. All three components of the burial assemblage were made of wood and decorated with hieroglyphic inscriptions and myriad vignettes from the *Book of the Dead* (for a review of the *Book of the Dead* during these dynasties, see Lenzo, 2023). The burial assemblages also included shabti figurines, papyri versions of the *Book of the Dead* and the *Amduat*, and other objects. For typological and chronological reviews of these so-called 'yellow coffins' in the context of the evolution of ancient Egyptian coffins as a whole, see Niwiński (1988) and Sousa (2018).

The variety of vignettes painted onto the coffins stems from the transition from the opulent tombs of the Ramesside period (19th and 20th Dynasties) to the simple, unadorned tombs of the 21st Dynasty. Niwiński (1988: 18) suggested that this transition stemmed from the need to defend the mummies from grave robbers. More importantly, Niwiński (1988: 18) argued that the paintings on the coffins of the 21st Dynasty were a continuation of the tomb decorations of the previous dynasties, though in a condensed form. In this context, we should think of the cosmological vignette as an evolution of the *Fundamentals* ... Geb, who is only mentioned in the text of the *Fundamentals* …, is now an integral part of the vignette. And while Nut is naked in most of the vignettes (Figure 1a), the stars that cover her body in some of the vignettes (Figure 1b) may have started off as a replacement of the list of decan stars that appeared along Nut's back in the Ramesside tombs. Some vignettes continued to depict Nut swallowing and giving birth to the Sun (Figure 1c), while others chose to depict her carrying the Sun on her back, either as a Solar disc, the Solar Bark, or as a scarab (Figure 1d–e). The latter may have been a way to merge Nut's descriptions in the *Fundamentals* … and the *Book of the Divine Cow* into one visual.

## 2 DATA

In order to study variations in Nut's appearance in the cosmological vignette, I began by searching for it on coffins on display at various museums around the world. At first, I relied on the digital collections that large museums, such as the Louvre and the British Museum, make available via their public-facing websites. I then expanded my search to include printed museum catalogs and scholarly publications that focused on the coffins of a specific museum or a specific dynasty. To be included in the sample, a coffin was required to have detailed photographs or sketches from all angles in order for me to determine whether it included a depiction of Nut.

At the time of writing, the catalog includes 555 coffin elements from all eras of ancient Egyptian history (Table 1). By coffin elements, I refer to sarcophagi, coffin cases, coffin lids, and mummy covers. During some eras, the deceased would be buried in a coffin set composed of several of these elements. As each museum presents its collection differently, some items in my catalog refer to a single element (e.g., a lid or a case) while others refer to part of an assemblage (e.g., an inner coffin case along with its lid). In this work, I do not include papyri, amulets, shabti figurines, and other objects that were buried together with the deceased. Due to its size, the full catalog is not included in this paper but will be shared on request.

Once it became clear that the majority of cosmological vignettes were found on coffin sets of the 21st and 22nd Dynasties, I expanded my search to include scholarly publications that focused on the coffin sets of this era. In this way, I assembled a sample of 75 cosmological vignettes. At this point, I was introduced to Andrzej Niwiński and Agnes Klische. By comparing their catalogs to mine, I grew my sample to 118 cosmological vignettes on 21st/22nd Dynasty coffins (Table 2) and five vignettes on coffins from later dynasties (Table 3). Thus, the sample presented in this paper reflects the current census of Nut's cosmological vignettes as they appear on ancient Egyptian coffins.

In the process of searching for examples of the cosmological vignette, I noted any depiction of Nut on the coffins in my catalog. It is important to note that, while Tables 2 and 3 provide a complete list of the known examples of cosmological vignettes on Egyptian coffins, the broader catalog from which they are derived is not complete. First, it does not represent an exhaustive search of all Egyptian collections. Second, due to the archival nature of this work, it is biased towards large museums whose collections have either been described in academic papers or digitized and made available on their public-facing websites. Third, only a fraction of the coffins in museum collections has been digitized from all angles. Fourth, because the large majority of cosmological vignettes is found on coffins from the 21st/22nd Dynasties, the catalog is biased towards coffins from that era. Despite these caveats, this sample does reveal certain trends in the way Nut was depicted on coffins throughout ancient Egyptian history (Section 5).

Throughout this paper, I refer to various coffins as they are numbered in Tables 2 and 3. Thus, the coffin of Nes-per-nebu in the collection of the Kunsthistorisches Museum in Vienna is referred to as C1. In most cases, I present the names of the coffins' occupiers as they appear in the publications I consulted. Egyptian names are transcribed differently in different languages, but the only times I attempted to standardize the transcription of specific names was when a coffin set was split between different museums or when a name from an older or non-English publication was updated by a later publication or one written in English. For more homogenous transcriptions, as well as the original hieroglyphs, see Niwiński (1988) and Klische (2024).

## 2.1 Notes on Individual Vignettes

Several of the cosmological vignettes in Table 2 deserve closer scrutiny. In two vignettes, namely C9 and C13, Nut was either described or sketched, respectively, as naked. However, images and sketches shared with me (A. Klische, pers. comm., 2024) revealed that Nut was, instead, covered in stars. In a similar vein, whereas I classified Nut as naked in C98, Agnes Klische classified her as covered in stars. My classification is based on an image of C98 from Bristol Museum & Art Gallery's digital collection. A close examination of a high-resolution version of this image reveals no traces of stars.

Table 1: Coffin sample, by Dynasty (following the chronology of Hornung, 1999).

| Dynasty | Years | Number |
|---|---|---|
| Unknown | — | 7 |
| Old Kingdom | | |
| 4 | 2600–2571 BCE | 1 |
| First Intermediate Period | | |
| 11 | 2081–1938 BCE | 4 |
| Unknown | 2198–1938 BCE | 3 |
| Middle Kingdom | | |
| 12 | 1938–1759 BCE | 9 |
| Second Intermediate Period | | |
| 17 | 1640–1539 BCE | 4 |
| New Kingdom | | |
| 18 | 1539–1292 BCE | 32 |
| 19 | 1292–1188 BCE | 8 |
| 20 | 1188–1075 BCE | 5 |
| 19/20 | | 1 |
| 20/21 | | 2 |
| Unknown | 1539–1075 BCE | 12 |
| Third Intermediate Period | | |
| 21 | 1075–945 BCE | 287 |
| 22 | 945–715 BCE | 49 |
| 23 | 818–715 BCE | 2 |
| 25 | 712–664 BCE | 42 |
| 22/23 | | 4 |
| 25/26 | | 20 |
| 22–26 | | 4 |
| Late Period | | |
| 26 | 664–525 BCE | 24 |
| 30 | 380–343 BCE | 7 |
| 26/27 | 645–404 BCE | 1 |
| Unknown | 664–343 BCE | 4 |
| Ptolemaic Period | | |
| Ptolemaic | 305–30 BCE | 13 |
| Roman Period | | |
| Roman | 30 BCE–642 CE | 10 |

In vignettes C7 and C110, Nut seems to be covered in spots instead of the more standard five-pointed stars. I treat these as simply a different depiction of stars. An exception to this rule is C77, in which Nut is mostly naked

Table 2: 21st/22nd Dynasty coffins with cosmological vignettes of Nut.

| No. | Museum | Country | Museum ID | Name* | Source |
|---|---|---|---|---|---|
| C1 | Kunsthistoriches Museum Wien | Austria | ÄS 6270 | Nes-per-nebu | Digital collection; Egner and Haslauer (2009); Virey (1910) |
| C2 | Museu Nacional, Rio de Janeiro | Brazil | Inv. 525-526 | Hori | Kitchen (1990: 133, Plate 121) |
| C3 | Metternich Museum, Lázně Kynžvart | Czech Republic | 1086 | Ken Amon | Digital collection |
| C4 | Graeco-Roman Museum, Alexandria | Egypt | 1297 | Khonsumes | Klische (2024: W1) |
| C5 | Alexandria National Museum | Egypt | 135 (Cairo J29670) | Akhaneferamon (O) | Niwiński (2023: 999) |
| C6 | Alexandria National Museum | Egypt | 136 (Cairo J29670) | Akhaneferamon (I) | Niwiński (2023: 1027) |
| C7 | Egyptian Museum in Cairo | Egypt | A.138 | Tashedkhonsu | Klische (2024: W18) |
| C8 | Egyptian Museum in Cairo | Egypt | CG 6002; J29658 | (Henttaui) | Chassinat (1909: 5); Klische (2024: W19) |
| C9 | Egyptian Museum in Cairo | Egypt | CG 6005; J29717 | Istemkheb (O) | Chassinat (1909: 15); Klische (2024: W20) |
| C10 | Egyptian Museum in Cairo | Egypt | CG 6008; J29740 | Istemkheb (I) | Chassinat (1909: Figure 26) |
| C11 | Egyptian Museum in Cairo | Egypt | CG 6011; J29718 | Pameshm | Klische (2024: W21) |
| C12 | Egyptian Museum in Cairo | Egypt | CG 6022; J29711 | 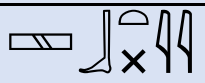 | Chassinat (1909: 73) |
| C13 | Egyptian Museum in Cairo | Egypt | CG 6038; J29702 | Tjanefer | Niwiński (1995: Figure 1); Klische (2024: W22) |
| C14 | Egyptian Museum in Cairo | Egypt | CG 6042; J29680 | Anonymous | Niwiński (1995: Figure 7) |
| C15 | Egyptian Museum in Cairo | Egypt | CG 6076; J29665 | Khaas | Niwiński (1995: Figure 85) |
| C16 | Egyptian Museum in Cairo | Egypt | CG 6079; J29668 | Paduamun (I) | Niwiński (1999: Figure 25); Abbas (2014: 53) |
| C17 | Egyptian Museum in Cairo | Egypt | CG 6081; J29668 | Paduamun (O) | Niwiński (1999: figure 13); Abbas (2014: 29) |
| C18 | Egyptian Museum in Cairo | Egypt | CG 6090; J29679 | Djedmutiusankh | Niwiński (2011: Figure 34) |
| C19 | Egyptian Museum in Cairo | Egypt | CG 6099; J29692 | Ankhefenmut | Niwiński (2023: 1389) |
| C20 | Egyptian Museum in Cairo | Egypt | CG 6130; J29625 | Tashedkhonsu | Klische (2024: W30) |
| C21 | Egyptian Museum in Cairo | Egypt | CG 6148; J29675 | Ankhensenmut | Klische (2024: W33) |
| C22 | Egyptian Museum in Cairo | Egypt | CG 6153; J29706 | Amunpermut | Niwiński (1999: Figure 132); Guilhou (2017) |
| C23 | Egyptian Museum in Cairo | Egypt | CG 6157; J29616 | Paifudjar | Klische (2024: W35) |
| C24 | Egyptian Museum in Cairo | Egypt | CG 6169; J29664 | Nespaneferher | Klische (2024: W36) |
| C25 | Egyptian Museum in Cairo | Egypt | CG 6173; J29649 | Amenniutnakht | Sousa (2020: 106, 128) |
| C26 | Egyptian Museum in Cairo | Egypt | CG 6180; J29659 | Nesyamenemope | Niwiński (1999: 64) |
| C27 | Egyptian Museum in Cairo | Egypt | CG 6183; J29660 | Djedmaatiusankh (I) | Niwiński (1999: Figure 58); Guilhou (2017) |
| C28 | Egyptian Museum in Cairo | Egypt | CG 6186; J29661 | Userhetmes (I) | Klische (2024: W40) |
| C29 | Egyptian Museum in Cairo | Egypt | CG 6188; J29662 | Tchent-ipet (O) | Niwiński (1999: Figure 108) |
| C30 | Egyptian Museum in Cairo | Egypt | CG 6190; J29662 | Tchent-Ipet (I) | Niwiński (1999: Figure 124); Guilhou (2017) |
| C31 | Egyptian Museum in Cairo | Egypt | CG 6201; J29678 | Shedsuamon (O) | Niwiński (2023: 1641) |
| C32 | Egyptian Museum in Cairo | Egypt | CG 6203; J29678 | Shedsuamon (I) | Niwiński (2011: Figure 35) |
| C33 | Egyptian Museum in Cairo | Egypt | CG 6209; J29663 | Ankhefenkhonsu | Klische (2024: W45) |
| C34 | Egyptian Museum in Cairo | Egypt | CG 6212; J29661 | Userhetmes (O) | Klische (2024: W46) |
| C35 | Egyptian Museum in Cairo | Egypt | CG 6214; J29660 | Djedmaatiusankh (O) | Niwiński (1999: Figure 38) |
| C36 | Egyptian Museum in Cairo | Egypt | CG 6229; J29667 | Mashasekebt | Klische (2024: W48) |
| C37 | Egyptian Museum in Cairo | Egypt | CG 6233; J29666 | Paduamun | Bruyère (1930: Figure 104) |
| C38 | Egyptian Museum in Cairo | Egypt | CG 6253; J29682 | Tjanefer | Klische (2024: W50) |
| C39 | Egyptian Museum in Cairo | Egypt | CG 6261; J29736 | Tjanefer | Niwiński (2011: Figure 30) |
| C40 | Egyptian Museum in Cairo | Egypt | CG 6265; J29733 | Dukhonsuiry (I) | Niwiński (2023: 1281) |
| C41 | Egyptian Museum in Cairo | Egypt | CG 6267; J29733 | Dukhonsuiry (O) | Niwiński (2023: 1233) |
| C42 | Egyptian Museum in Cairo | Egypt | CG 6272; J29628 | Menkheperre | Klische (2024: W54) |
| C43 | Egyptian Museum in Cairo | Egypt | CG 6279; J29737 | Tairetra | Niwiński (2023: 1097) |
| C44 | Egyptian Museum in Cairo | Egypt | CG 6291; J29611 | Nesimen (O) | Klische (2024: W56) |
| C45 | Egyptian Museum in Cairo | Egypt | CG 6293; J29611 | Nesimen (I) | Klische (2024: W57) |
| C46 | Egyptian Museum in Cairo | Egypt | J29620 | Djedkhonsuiufankh (F) | Niwiński (2021: Figure 8), Klische (2024: W62) |
| C47 | Egyptian Museum in Cairo | Egypt | J29635 | Gatseshen | Niwiński (2023: 1604), Klische (2024: W63) |
| C48 | Egyptian Museum in Cairo | Egypt | CG 61027 | Masahirta | Daressy (1909: Plate 37) |
| C49 | Egyptian Museum in Cairo | Egypt | CG 61030 | Nesikhonsu | Daressy (1909: Plate 48) |
| C50 | Egyptian Museum in Cairo | Egypt | CG 61031 | Astemkheb | Daressy (1909: Plate 52) |
| C51 | Egyptian Museum in Cairo | Egypt | CG 61032 | Taiuhert | Daressy (1909: Plate 56) |
| C52 | Egyptian Museum in Cairo | Egypt | — | Anonymous (F) | Klische (2024: W64) |
| C53 | Egyptian Museum in Cairo | Egypt | — | Anonymous | Klische (2024: W65) |
| C54 | El Arish | Egypt | (Cairo J29651) | Sennu | Niwiński (2023: 953) |
| C55 | Louvre | France | 5171 | Bakenkhonsu | Digital collection |
| C56 | Louvre | France | E 10636 | Djedkhonsouioufânkh | Digital collection |
| C57 | Louvre | France | E 13034 | Tanetnahereret | Digital collection |
| C58 | Louvre | France | E 13036 | Anonymous woman | Niwiński and Rigault-Déon (2024: 326-327) |
| C59 | Louvre | France | N 2562 | Tanetimen (I) | Digital collection |
| C60 | Louvre | France | N 2570 | Paser (I) | Niwiński and |
| | | | | | Rigault-Déon (2024: 211) |
| C61 | Louvre | France | N 2581 | Paser (O) | Digital collection |
| C62 | Louvre | France | N 2610 | Soutymes | Digital collection |
| C63 | Musée d'Aquitaine, Bordeaux | France | Mesuret-8590 | Iteneferamon | Dautant et al. (2011); Perrot (1844) |
| C64 | Musée des Beaux-Arts et d'Archéologie, Besançon | France | A.778 | Seramon | Niwiński (2011: Figure 43) |
| C65 | Musée Testut Latarjet d'Anatomie de Lyon | France | — | Kamenishery | Jamen (2016); Dautant and Jamen (2017) |
| C66 | Musée d'Archéologie Méditerranéenne de Marseille | France | 5171 | Bakenkhonsu | Dautant (2013), Klische (2024: W79) |
| C67 | Musée Centre de la Vieille Charité, Marseille | France | 253.1 | Khonsumes | Klische (2024: W78) |
| C68 | Musée Departmental Dobreé, Nantes | France | 56.6424 | Ankhkheriretiu | Klische (2024: W81) |
| C69 | Museum of Natural History Perpignan | France | — | Iufenkhonsu | Niwiński (2023: 540) |
| C70 | Unknown museum in Paris (?) | France | — | — | Lanzone (1881: 407, Plate 158) |
| C71 | Ägyptisches Museum und Papyrussammlung | Germany | ÄM 8 | Tanetimen (O) | Klische (2024: W3) |
| C72 | Ägyptisches Museum und Papyrussammlung | Germany | ÄM 28 | Taiuheret | Klische (2024: W4) |
| C73 | Museum of Fine Arts Budapest | Hungary | 87.5-E | Horhotep (F) | Kóthay and Liptay (2013: 84-85) |
| C74 | Museum of Fine Arts Budapest | Hungary | 51.325 | Unknown (F) | Digital collection |
| C75 | Museo Egizio | Italy | Cat. 2214 | — | Niwiński (2011: Figure 45); S. Töpfer (pers. comm., 2024) |
| C76 | Museo Egizio | Italy | Cat. 2226 | Tabakenkhonsu | Digital collection |
| C77 | Museo Egizio | Italy | Cat. 2228 | Tamutmutef | Digital collection |
| C78 | Museo Egizio | Italy | Cat. 2236 | Butehamon | Digital collection |
| C79 | Museo Egizio | Italy | Cat. 2237 | Butehamon | Digital collection |
| C80 | Museo Egizio | Italy | Cat. 2238 | Khonsumes | Digital collection |
| C81 | Museo Archeologico Nazionale di Firenze | Italy | 8527 | Khonsumes | Sousa (2019: 183) |
| C82 | Museo Archeologico Nazionale di Firenze | Italy | 8528 | Djedmutiuesankh | Sousa (2019: 101) |
| C83 | Civico Museo d'Antichità J.J. Win-ckelmann, Trieste | Italy | 12086 | Anonymous woman | Digital collection |
| C84 | Rijksmuseum van Oudheden | Netherlands | F 93/10.1a | Gautseshen (O) | Digital collection |
| C85 | Rijksmuseum van Oudheden | Netherlands | F 93/10.1b | Gautseshen (I) | Digital collection |
| C86 | Rijksmuseum van Oudheden | Netherlands | AMM 2-a | Amenhotep | Digital collection |
| C87 | Rijksmuseum van Oudheden | Netherlands | AMM 18-h | Djedmontefanch | Digital collection |
| C88 | Pushkin State Museum of Fine Arts | Russia | I.1.a 6800 | Iusankh | Lavrentyeva (2021) |
| C89 | National Museum of the Tatarstan Republic | Russia | КП-5404/1 | Nesitaudjatakhet (I) | Bolshakov (2017); Turaev (1904, 1914) |
| C90 | National Museum of Warsaw | Poland | MNW 139072 | Anonymous | Bruyère and Bataille (1936: Plate 6) |
| C91 | Geographical Society of Lisbon | Portugal | SGL-AC-519 | Henut-taui | Sousa (2017: Plates 83, 89) |
| C92 | Museo Arqueologico Nacional | Spain | 18257 | Asetemakhbit | Digital collection |
| C93 | Museo Arqueologico Nacional | Spain | 15216 | Amenemhat | Digital collection |
| C94 | Museum Gustavianum | Sweden | VM 228 | Khonsumes | Digital collection |
| C95 | Medelhavsmuseet | Sweden | MME 1977:018 | Unknown (F) | Digital collection |
| C96 | Musée d'Art et d'Histoire, Geneva | Switzerland | A 2016-0014 | Tanetnakht (F) | Digital collection |
| C97 | Bolton City Museum | UK | 69.30 | Taiuhenut | Klische (2024: W6) |
| C98 | Bristol Museum & Art Gallery | UK | Ha7386.1037 | Horemkenesi | Digital collection; Dawson et al. (2002: 43); Taylor (1996: Plate 4) |
| C99 | The British Museum | UK | EA15658 | Amenhotep (F) | Klische (2024: W72) |
| C100 | The British Museum | UK | EA29591 | Inpehefnakht | Digital collection |
| C101 | Fitzwilliam Museum | UK | E.1.1822 | Nespawershefyt (O) | Digital collection; in-person visit |
| C102 | Fitzwilliam Museum | UK | E.1.1822 | Nespawershefyt (I) | Digital collection; in-person visit |
| C103 | Fitzwilliam Museum | UK | E.1.1822 | Nespawershefyt (M) | Digital collection; in-person visit |
| C104 | National Museums Scotland | UK | A.1907.569 | Iufenamun | Manley and Dodson (2010: 49) |
| C105 | Swansea University Egypt Centre | UK | W1982 | Iwesenhesetmut | Digital collection |
| C106 | Odessa Archaeological Museum | Ukraine | OAM 71695 | Nesimut | Tarasenko (2017) |
| C107 | Odessa Archaeological Museum | Ukraine | OAM 52976 | Nesitaudjatakhet (O) | Tarasenko (2016) |
| C108 | Boston Museum of Fine Arts | USA | 54640 | Henuttaui | Klische (2024: W8) |
| C109 | Brooklyn Museum | USA | 08.480.2a-c | Pasebakhaienipet | Digital collection |
| C110 | Carnegie Museum of Natural History, Pittsburgh | USA | 1-1 | Anonymous | Niwiński (2023: 541) |
| C111 | Carnegie Museum of Natural History, Pittsburgh | USA | 22266-3 | Anonymous | Niwiński (2023: 542) |
| C112 | Los Angeles County Museum of Art | USA | M.47.3a-c | Anonymous | Digital collection |
| C113 | The Metropolitan Museum of Art | USA | 25.3.8b | Menkheperre | Klische (2024: W84) |
| C114 | The Metropolitan Museum of Art | USA | 25.3.183 | Henettawy | Klische (2024: W83) |
| C115 | The Metropolitan Museum of Art | USA | 30.3.24a, b | Nauny | Digital collection; in-person visit |
| C116 | Michael C. Carlos Museum | USA | 1999.001.017A | Tanakhtnettahat | Digital collection |
| C117 | National Museum of Natural History | USA | A154954-0 | Tenetkhonsu | Digital collection |
| C118 | Virginia Museum of Fine Arts, Richmond | USA | L.7.47.72 | Itamun | Digital collection |

*Key: Rows with numbers in green highlighting denote vignettes in which Nut is decorated with stars.
In the 'Name' column, (O), (I), (M), and (F) denote outer coffin, inner coffin, mummy cover, and fragment, respectively.

except for a spotted rug on her back, on which rests a winged Solar disc (Figure 1e). A similar spotted rug is often seen in a different vignette on the back of the sacred cow as she emerges from the Western Mountain. On the C77 coffin, both the cow and its rug are spotted. Since the spots do not cover all of Nut's body in this vignette, I do not treat C77 as an example of a star-studded Nut.

Intriguingly, in vignettes C75 and C78, Nut is painted almost or entirely black, respectively

(Figure 1g–h). This might be a different way to depict Nut as the night sky. However, in C75, Geb is also painted black (though Shu is painted the usual brown), while in C78 Geb, Isis, and Nephthys are all painted black as well. Thus, it is unclear what exactly the black hue is supposed to symbolize in these vignettes.

Vignette C70 is only attested by Lanzone (1881: 407, Plate 158), who sketched Nut without stars. However, a very similar sketch appears in Wilkinson (2003: 129), in which stars cover Nut's torso (but not her arms or legs). Lanzone (1881) claimed to have seen the vignette in an unnamed museum in Paris (which is probably not the Louvre, since the latter museum is noted, by name, in the second half of the page when referring to a vignette of Nut on a papyrus in the Louvre's collection). Wilkinson (2003) does not provide a source for the sketch, only noting that it was created by Philip Winton. Both sketches are nearly identical to C117, in the collection of the Smithsonian Institution's National Museum of Natural History in Washington, D.C. An image of C117 from the museum's digital collection clearly shows Nut's torso covered in stars, as in Philip Winton's sketch. I suggest that C70 and C117 are two separate coffins. The whereabouts of the first, which was sketched by Lanzone (1881), have been lost. The second coffin was only discovered in 1891 as part of the second cache of Deir el-Bahari and has been in the Smithsonian's collection since it was gifted to the US by the Khedive of Egypt in 1893 (see Section 3, below). It is most likely that Philip Winton sketched the latter coffin.

Vignette C75, which was not included in the Niwiński (1988) catalog, is shown in Figure 45 in Niwiński (2011). However, Niwiński (2011) does not name the owner of the coffin. The Museo Egizio's record of this coffin also lacks a name, as the museum does not hold the coffin's lid in its collection (S. Töpfer, pers. comm., 2024). Hence, this coffin remains unnamed in Table 2. The same is true of C70, which is not named by Lanzone (1881).

Vignette C86 is a depiction of Nut in her arched form, covered in stars, painted on the interior of the headboard of the coffin case. Both Shu and Geb are absent. These properties mark C86 either as a peculiar version of the cosmological vignette (such as C103, in which Nut, held by Heka, is painted on the inside of a mummy cover) or as an independent depiction of Nut. Since this is the only instance in which I have seen Nut depicted in her arched, starred form on a 21st/22nd Dynasty coffin outside of the standard cosmological vignette, I have elected to include it in my sample.

Finally, the cataloging of vignettes C7 and C12 is unclear. The coffins in the collection of the Egyptian Museum in Cairo have been cataloged by several Egyptologists over a span of nearly a century (Chassinat, 1909; Daressy, 1909; Moret, 1913; Niwiński, 1995; 1999), and the work is still far from finished. These coffins are cataloged using three different numbers. Coffin sets are given a journal number (abbreviated as J, JE, or JdE, depending on the publication), while individual coffin elements are given a Catalogue Général (CG) number. For example, the outer and inner coffins of Paduamun are cataloged as CG 6079 and CG 6081, respectively, but they both belong to J29668. Coffin sets from the second cache from Deir el-Bahari (also known as the Bab el-Gasus cache; see Section 3, below) were first cataloged by Daressy (1907) with A numbers. Paduamun, for example, is A.87. Since only a fraction of the vignettes in Table 2 belong to the second cache from Deir el-Bahari, their Daressy numbers are not noted here (they are, however, noted in the full catalog, available by request). In most cases, the J, CG, and A numbers of a specific coffin element are in agreement between different publications. The two exceptions are C7, which Klische (2024) only cataloged according to the deceased's Daressy number, and C12, originally cataloged by Chassinat (1909: 70) as CG 6022. Niwiński (1988: 129) associates the latter coffin element with J29711 and A.86, which has different CG numbers (6027–6029). Here, I note Chassinat's original CG number along with the J number provided by Niwiński (1988). Neither Chassinat nor Niwiński provide a transcription of the owner's name, so I have only reproduced its hieroglyphic form.

Table 3: Coffins with cosmological vignettes of Nut from later Dynasties.

| No. | Museum | Country | Museum ID | Name | Dynasty | Source |
|---|---|---|---|---|---|---|
| C119 | Egyptian Museum in Cairo | Egypt | CG 41003 | Neskhonsout II | 26 | Moret (1913: Plate 11) |
| C120 | Egyptian Museum in Cairo | Egypt | TR 21.10.40.1 | Ankhrui | 30 | Petrie (1889: Plate 2) |
| C121 | Louvre | France | E 3913 | Tachepenkhonsou | 25 | Digital collection |
| C122 | The British Museum | UK | EA15655 | Hor | 25 | Digital collection |
| C123 | Unknown | ? | — | Ir | 26 | Sharpe and Bonomi (1858: Plate 1) |

## 3 THE COSMOLOGICAL VIGNETTE ON THE OUTER COFFIN OF NESITAUDJATAKHET

In 1891, Eugène Grébaut and Georges Daressy unearthed the tomb of 153 priests and priestesses of Amun from the 21st Dynasty in Deir el-Bahari, next to the first courtyard of the temple of Queen Hatshepsut (Daressy, 1900). The tomb included more than 200 coffins (Daressy, 1907) and their associated papyrus scrolls and other funerary artefacts. Most of the coffins from what has come to be known as the Bab el-Gasus (Gate of the Priests) Cache were retained for the Giza Museum, now the Egyptian Museum in Cairo. In 1893, a fraction of the coffins was divided into 17 lots that were gifted by the Khedive of Egypt, Abbas II Hilmy (عباس حلمي باشا), to (by lot number): France, Austria, Turkey, the UK, Italy, Russia, Portugal, Switzerland, USA, Netherlands, Greece, Spain, Sweden and Norway, Denmark, and the Vatican (for reviews surrounding the history of the Bab el-Gasus Cache and the current whereabouts of its coffins, see, e.g., Amenta and Guichard, 2017; Sousa et al., 2021).

Lot 6, which was gifted to what was then the Russian Empire, arrived in Odessa in 1894. From there, the various artefacts, which included five coffins, were sent to museums throughout Russia. Here, I will focus on the coffin set of the priestess Nesitaudjatakhet ( *nsy-t3-wḏ3t-3ḫt*), whose titles included 'mistress of the house', 'chantress of Amun-Re', and 'musician of Mut'. Numbered A.96 by Daressy (1907), the burial assemblage of Nesitaudjatakhet was divided between several museums (Tarasenko, 2016; 2017; 2019). The outer coffin remained in Odessa (C107) but the inner coffin was sent to Kazan (C89). Her mummy board is thought to be kept in the Ivan Kramskoi Museum of Fine Arts in Voronezh, while the rest of her burial assemblage is spread throughout the world; her funerary papyri are in the Egyptian Museum in Cairo and her shabti box is in the collection of the St. Gallen Historisches und Völkerkundemuseum in Switzerland.

Bolshakov (2017) has published a complete description of Nesitaudjatakhet's inner coffin, and a complete description of the outer coffin is in preparation (M. Tarasenko, pers. comm., 2024). The existing papers show that both sets of coffins are adorned with Nut's cosmological vignette. The cosmological vignette on the inner coffin is unremarkable (Figure 1c). Nut is naked and arched over Geb, who is covered in reeds. One winged Solar disc is painted close to her mouth, representing the Sun being swallowed as it sets, while a second winged Solar disc is painted next to her groin, representing the Sun being reborn as it rises in the east. Shu is absent, his place taken by the Solar Bark ferrying Re and a winged uraeus.

The cosmological vignette on Nesitaudjatakhet's outer coffin (Figure 2) deserves a more detailed description. Nut appears naked, save for an ankle bracelet, and her body is covered with stars and an undulating black curve, discussed in detail below. Nut is flanked by winged uraei with urea rings ( *šn*) between their outstretched wings. The hieroglyphs surrounding the winged uraeus next to Nut's head ( *nỉt*) indicate it is a depiction of the Goddess Neith. It is unclear whether the winged uraeus on Nut's other side is also linked to Neith, as it lacks the hieroglyph . Below the uraei, Nut's name is written as  along with a Solar disc ( ), ankh ( *ꜥnḫ*), and a *was* scepter ( *w3s*). The *was* scepters may symbolize the pillars of the sky (M. Tarasenko, pers. comm., 2024), while the Solar disc may represent Nut's role in the Solar cycle. Nut is held by Shu, who also holds the determinative for the word sky ( *pt*) on which rests a boat, probably the Solar Bark. The hieroglyphs between Shu's arms describe him as "the great god, lord of the sky" ( *nṯr ꜥ3 nb pt*). The string of hieroglyphs on Shu's right side reads: "Osiris, Foremost of the West, Neith" ( *wsỉr ḫnt ỉmnt nỉt*), while the hieroglyphs next to Geb's head read "giving offerings" ( *dỉ ḥtp ḥrt3*). It may be that both sets of hieroglyphs combine to form one phrase, as a similar phrase invoking offerings for Neith appears as part of a longer inscription along the right side of the inner coffin (Bolshakov, 2017: 193). On either side of Shu, two falcons hold ankhs and *was* scepters. Geb, covered in green reeds, reclines at Nut's feet. Next to Geb is a long, coiled snake, which may indicate Geb's connection to snakes, as described in the *Book of the Divine Cow*. Similar snakes appear next to Geb in vignettes C33, C67, C93, C117, and C118.

Neith's presence in this vignette, if only in hieroglyphic form, is intriguing, as it may be that Nut and Neith are being conflated. Although the two Goddesses developed separately (Neith as a Hunter/Warrior Goddess known as 'mistress of the bow' and 'ruler of arrows', and Nut as the Goddess of the Sky), they both evolved to embody additional cosmological, maternal, and eschatological roles (Hol-lis, 2019: 7–29; Sethe, 1906: 144–147; Wilkinson, 2003: 156–159), and Nut and Neith also often fulfill the same roles as the coffin and as guardians of the deceased. Assmann (2005: 165) provides an example of this conflation in the texts carved into the 19th Dynasty sarcophagus of King Merneptah, and I have seen visual examples of such conflations in the coffins in my sample (Figure 3). For example, Neith is sometimes interposed instead of Nut as the kneeling goddess with outstretched wings on coffin lids (as on the lid of Nesitaudjatakhet's inner coffin; Bolshakov, 2017: Figure 1); as the principal standing figure inside the coffin (e.g., the mummy cover of Mashasekebt, Egyptian Museum in Cairo CG 6238; Niwiński, 1988: Plate 15a); or as the Tree Goddess who nourishes the deceased (e.g., Niwiński, 1995: Figure 14). Moreover, winged uraei bearing Neith's name are seen in various compositions on coffins, e.g., on the lids of certain coffins, where a pair of winged Neith uraei flank an image of a kneeling, winged Nut (e.g., Niwiński, 1999: Plate 1.2, Plate 6.1). Among the cosmological vignettes in my sample, Neith's name also appears above winged uraei in vignettes C4, C19, C34, C71, and C110. In other vignettes, the winged uraei bear the names of Isis and Nephthys, raising the possibility that Neith plays here the same guardian role attributed to her in the *Pyramid Texts* along with Isis, Nephthys, and Serket, as in spell 362:

> My father! My father in the darkness! My father, Atum in the darkness! Fetch me to your side, and I will light a lamp for you and guard you like Nut's guarding of those four serpent-goddesses on the day they guarded the chair—Isis, Nephthys, Neith, and She who Aspirates Throats. (Allen, 2015: 81).

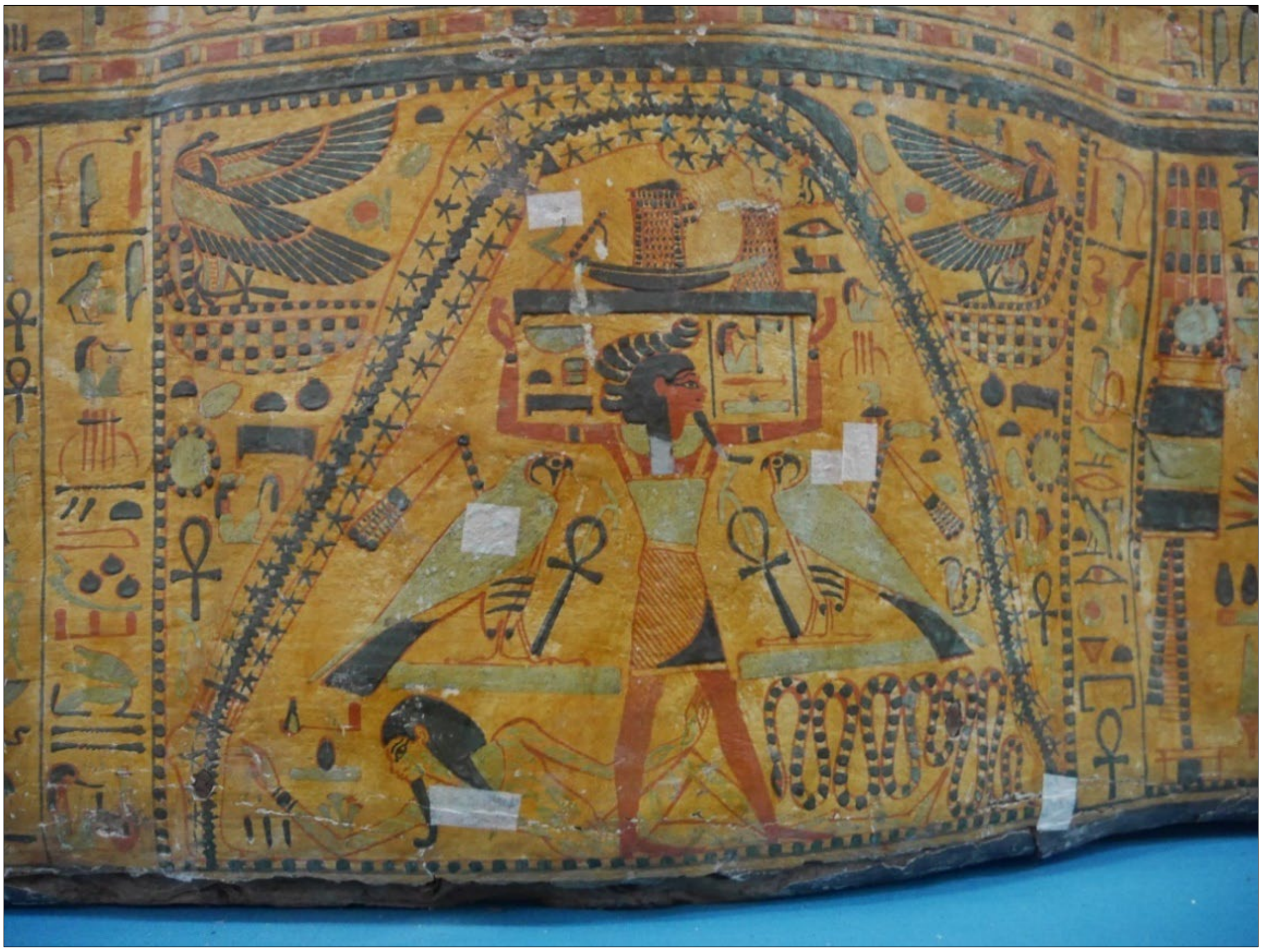

Figure 2: Nut's cosmological vignette on the outer coffin of Nesitaudjatakhet in the collection of Odessa Archaeological Museum OAM 52976 (C107). Nut's body is covered in stars as well as a thick, undulating black curve that runs from the soles of her feet to the tips of her fingers. This curve, surrounded by stars on both sides, is reminiscent of the Milky Way's Great Rift (courtesy: Photo by Mykola Tarasenko; © Odessa Archaeological Museum).

## 3.1 A Visual Depiction of the Milky Way

The most peculiar feature of Nut's cosmological vignette in Figure 2 is the undulating, thick black curve that crosses her body from the soles of her feet to the tips of her fingers. This curve runs down the middle of Nut's body, with the stars painted in roughly equal number above and below it. This curve is not an artistic technique to separate Nut's legs from one another, as that is achieved with a thinner curve that is half obscured by the stars below the undulating curve. A similar thin curve also separates Nut's arms. To further accentuate the undulating curve, it is drawn in black, while the thinner curves that separate Nut's limbs are drawn using the same red paint that is used to draw her outline. There is no written description of this curve in the scene itself. A complete description of the outer coffin is needed to ascertain whether such a description may be found in the texts that surround the vignette, but this is unlikely; most of the texts that adorn yellow coffins are rote invocations for the protection of various deities (see, e.g., the translations of the texts on the inner coffin of Nesitaudjatakhet; Bolshakov, 2017).

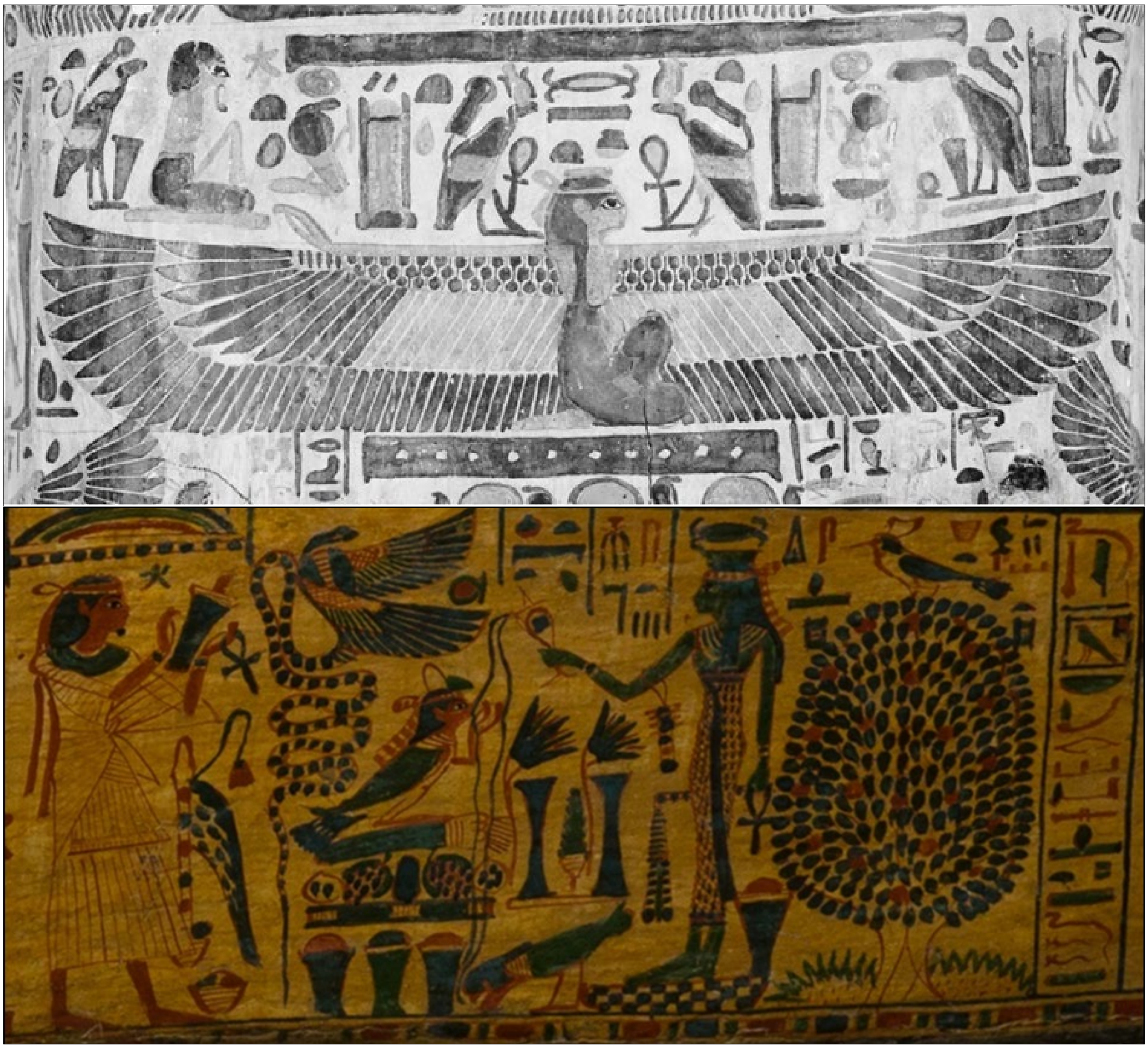

Figure 3: Examples of conflation between Nut ( ) and Neith ( ) on (top) the coffin lid of Djedkhonsouiouefânkh (courtesy: Louvre E10636; © 1980 Musée du Louvre, Dist. GrandPalaisRmn / Maurice et Pierre Chuzeville) and (bottom) the coffin of Khonsumes (courtesy: Museum Gustavianum VM 228).

Based solely on the visual aspects of the vignette, I offer two possible meanings for the nature of this curve. The first is that it represents the division of the sky in two, as noted in the *Pyramid Texts*, e.g., in spell 274 (Allen, 2015: 55): "… both skies go around (in service) for me and the two shores serve me." or in spell 332 (Allen, 2015: 73): "… The two skies have gone to me, the two lands have come to me. …" The second possibility is that the undulating curve represents the Milky Way. In this scenario, the undulating curve could be a representation of the Great Rift, the dark band of dust that cuts through the Milky Way's bright band of diffuse light. A comparison with a photograph of the Milky Way (Figure 4) shows the stark similarity between the undulating curve running through Nut's body and the un-dulating curve created by the dark nebulae that make up the Great Rift. Together with Nut's arched form, the undulating curve accentuates the apparent similarity between Nut's depiction and the appearance of the Milky Way in the night sky.

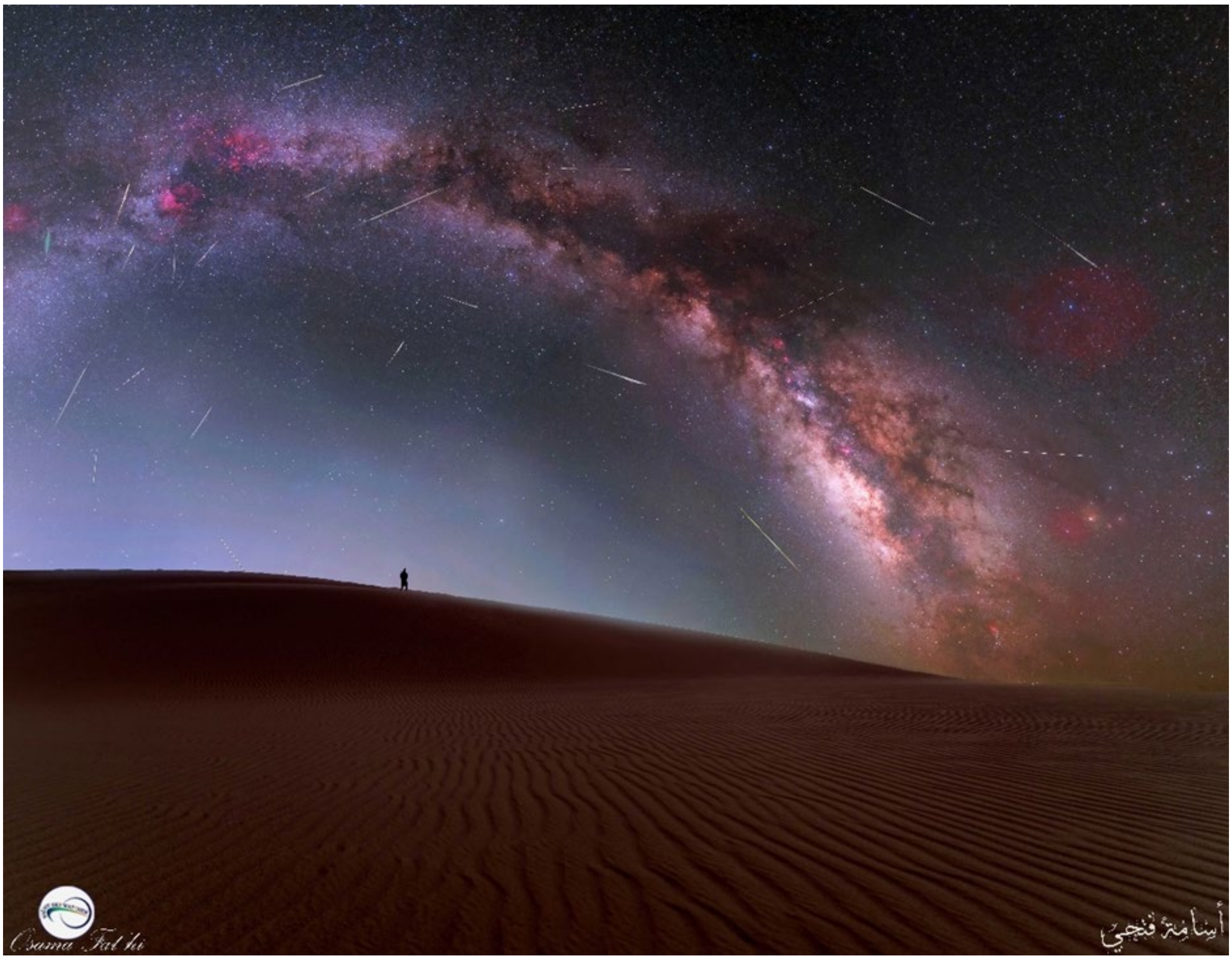

Figure 4: The Milky Way over the sand dunes of the Egyptian Western Desert near El-Fayoum, taken on 5–7 August 2022 at 12:00 AM (total exposure: 1.2 hours). Note the similarity between the Great Rift and the undulating black curve that bisects Nut's body in Figure 2 (courtesy: © Osama Fathi).

In one form or another, Nut is depicted on 312 (56%) of the coffin elements in my sample. Yet I have seen no other depiction of Nut that included a similar undulating curve, whether in her winged forms, full-length portraits, or cosmological depictions. The rarity of this curve reinforces the conclusion reached by Graur (2024a): Nut and the Milky Way may be linked, but they are not one and the same. We should not think of the Milky Way as a representation of Nut. Instead, we should think of the Milky Way as one more astronomical phenomenon that, like the stars and the Sun, are part of the sky and hence a part of Nut. Spell 69 of the *Book of the Dead* notes (Faulkner, 1972: 70): "I am Orion who treads his land, who precedes the stars of the sky which are on the body of my mother Nut …" It is in this light that we should understand Nut's depictions in the cosmological vignettes: Nut is the entire sky, and all celestial phenomena—the Sun, the stars, the Milky Way—are but markings on her body.

## 3.2 The Undulating Curve in New Kingdom Tomb Murals and a Link to the Winding Waterway

Similar undulating curves appear in four tombs in the Valley of the Kings. In the tomb of Seti I (KV 17), the astronomical ceiling in burial chamber J is split into the northern sky, with depictions of the circumpolar constellations, and the southern sky, with depictions of the planets, decans, and southern constellations. Each sky is bordered by two series of golden half circles arranged in such a way that a dark undulating curve takes form between them (Figure 5). A similar version of these borders appears in the tomb of Tausert and Setnakht (KV 14) on the ceiling of burial chamber J1. In the tomb of Ramesses IV (KV 02), the ceiling of burial chamber J is split between *Fundamentals …* and the *Book of the Night*. Both books include arched figures of Nut. These are displayed back-to-back and separated by a thick, golden undulating curve within two similarly colored straight lines. The ceiling of burial chamber J in the tomb of Ramesses IX (KV 06) also features two images of Nut as part of the *Books of the Day and the Night*. Here, the two depictions of Nut are separated by a zigzagging curve made up of several thin black lines.

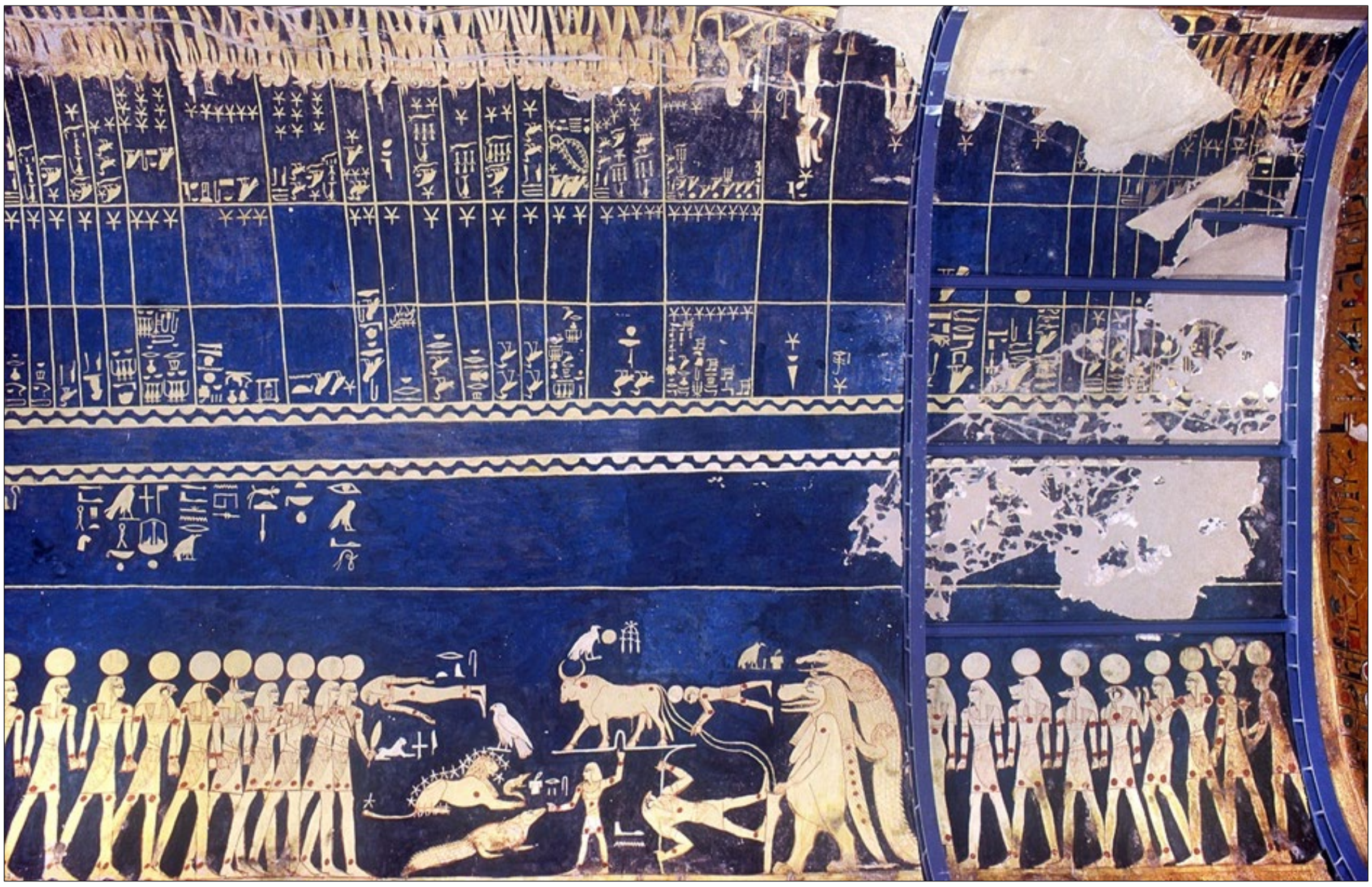

Figure 5: The astronomical ceiling from the tomb of Seti I (KV 17). Note the undulating black curves between rows of yellow half-circles that border the two halves of the ceiling (courtesy: Theban Mapping Project, Photographer, Francis Dzikowski, May 2000).

The most important example of the undulating curve appears in the tomb of Ramesses VI (KV 09). Here, the ceiling of burial chamber J is split between the *Book of the Day* and the *Book of the Night* (Figure 6). On each side, the books are bordered by an arched depiction of Nut, so that the two depictions of the goddess appear back-to-back. In both depictions, Nut's body is studded with stars. The *Book of the Day* also includes multiple Sun discs that represent the Sun's movement across the sky. Two undulating yellow curves separate the two representations of Nut. In each of the figures, the undulating curve begins at the base of Nut's hair, runs along (but above) her back, and finally curves over her rear. The connection of the curve to Nut's hair and rear shows that it is directly connected to her and not a mere decorative border. Together, the appearances of the undulating curves described here reinforce my suggestion that the curve bisecting Nut's body in the cosmological vignette is the Milky Way and that it splits the sky in two.

Identifying these undulating curves as depictions of the Milky Way also reopens the possibility that (*mr-nḫ3*), translated as "Winding Waterway," "Winding Canal," or "Shifting Waterway" was the ancient Egyptian name for our Galaxy. This link was first suggested by Davis (1985) but countered by Krauss (1997: 63–64), who argued for its identification with the ecliptic. Consider, however, the following spells from the *Pyramid Texts* in the context of the appearance of the undulating curve along Nut's body.

> The Nurse Lake is opened up, the Winding Canal becomes inundated, the Fields of Reeds have filled, so that I might be truly ferried to that eastern side of the sky, to the place where the gods are born, and I am born there in my birth with them as Horus, as him of the Akhet.
> PT 265 (Allen, 2015: 130).

> So, you go away to the sky, for the paths of the (sky's) arcs that ascend to Horus are swept for you. The mind of Seth is fraternal toward you as the great one of Heliopolis, when you have traveled the Winding Canal in Nut's north as a star that crosses the Great Green that is under Nut's belly. The Duat lays down your hand toward the

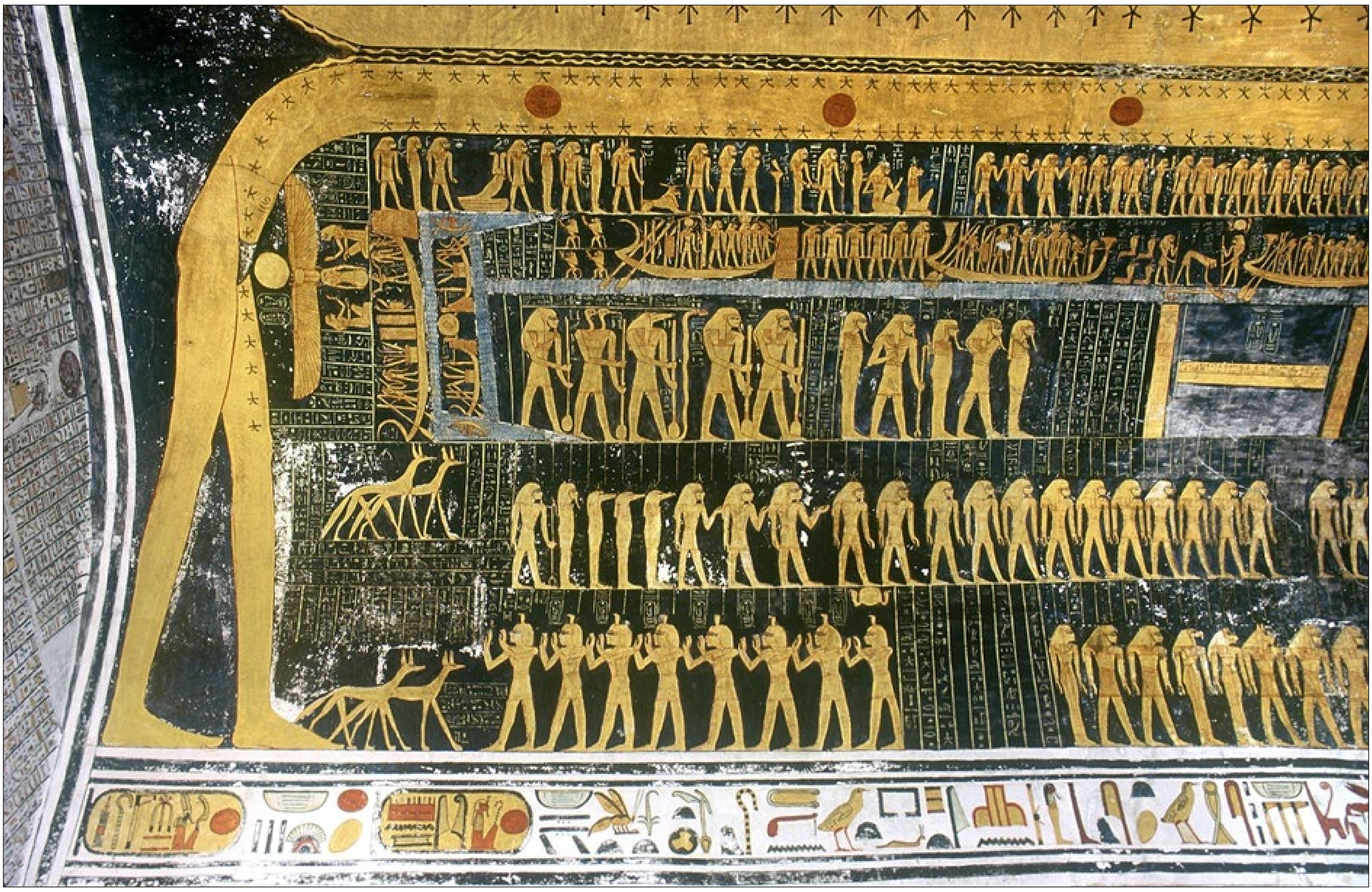

Figure 6 (above): The *Book of the Day* from the tomb of Ramesses VI (KV 09). Note the undulating yellow curve that runs above Nut's back, from her head to her rear. An identical curve runs along the back of the second depiction of Nut in the *Book of the Night* (courtesy: Theban Mapping Project, Photographer, Francis Dzikowski, November 1999).

place where Orion is, the sky's bull having given you his arm.
PT 437 (Allen, 2015: 110).

You go away to the sky on your metal chair and cross the Winding Canal, your face in the north of Nut. The Sun calls for you from the sky's zenith, and you ascend to the god.
PT 483 (Allen, 2015: 137).

I have gone up on Shu, I have climbed on the wing of Evolver. Nut is the one who has received my arm, Nut is the one who made a path for me.
Greetings, you two falcons in the prow of that boat of the Sun, who sail the Sun to the east! May you lift me and raise me up to the Winding Canal. When you put me among those imperishable gods and I make landfall among them, I cannot perish, I cannot end.
PT 624 (Allen, 2015: 241).

On their own, these spells (and others not quoted here) do not provide conclusive evidence for identifying the Winding Waterway with the Milky Way. However, once coupled with the winding appearance of the undulating curve and its connection to Nut, it becomes possible to imagine the Galaxy's Great Rift as the Winding Waterway along which the Sun and the deceased king are ferried from west to east before both are reborn at dawn (as opposed to the ecliptic, along which the planets move from east to west). This proposal is strengthened by the astronomical ceiling in the tomb of Seti I, where the undulating curve, like the Milky Way, separates the two skies instead of appearing solely in conjunction with the planets in the southern sky, as one would expect of the ecliptic.

## 4 A CROSS-CULTURAL ANALYSIS

An examination of the ways in which the Milky Way is visually depicted by other cultures strengthens the identification of the undulating curve in Figures 2 and 6 with the Milky Way. Several authors (e.g., Berezkin, 2010; Graur, 2024b; Krupp, 1991; 1995; Lebeuf, 1996) have noted that different cultures around the world have conceptualized the Milky Way in similar terms, even when those cultures were spatially and temporally disconnected. Many cultures, for example, view the Milky Way as a conduit between this life and the next. The Winding Waterway, which might be the Milky Way's Egyptian name (see above), fulfills the same role, ferrying the deceased king across the sky to the east, where he is reborn as one of the imperishable stars.

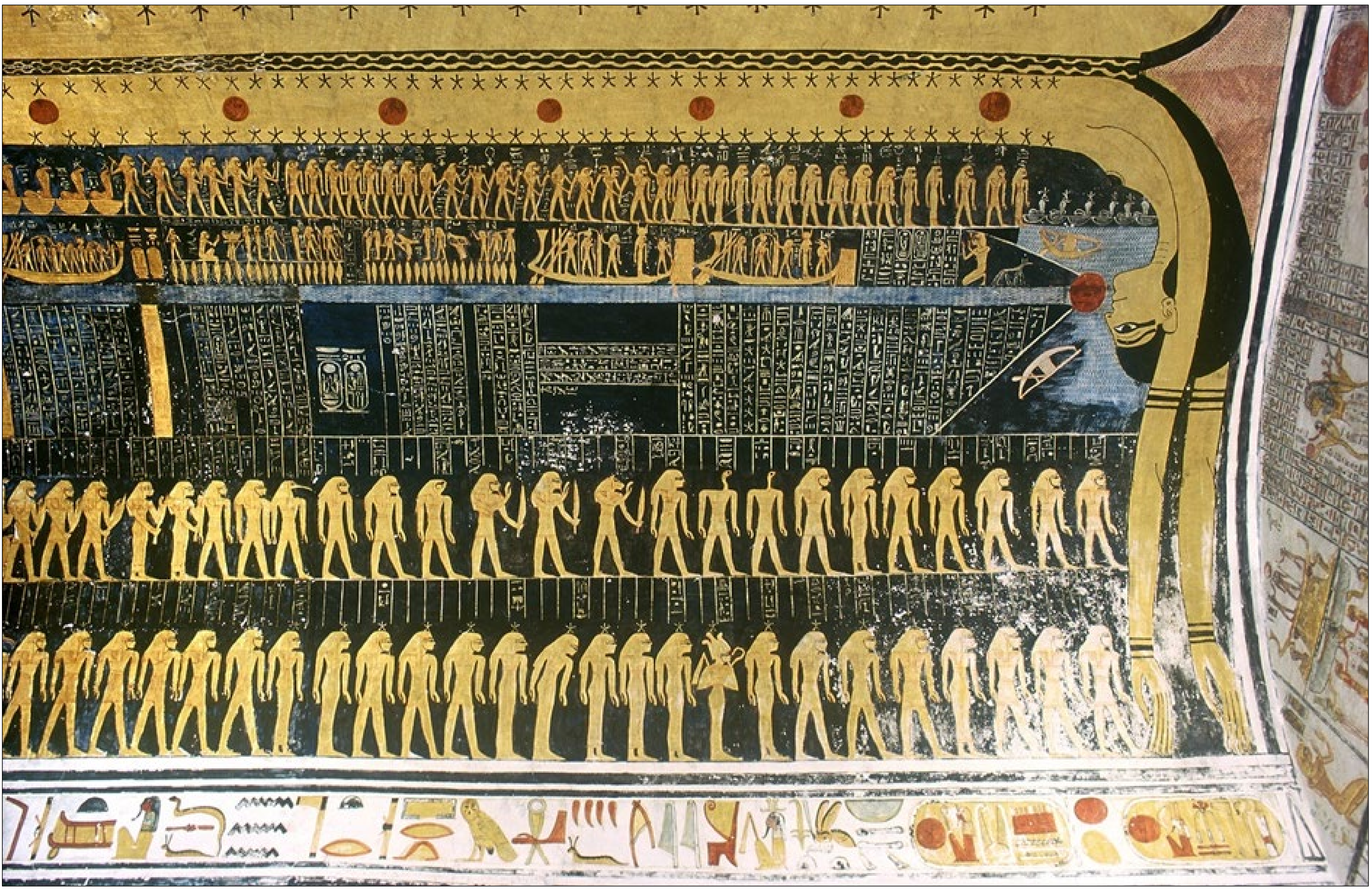

Some cultures have conceptualized the Milky Way as a deity or as a feature of a sky deity. Graur (2024a: 37, Figure 5) compared Nut to the Mesoamerican Goddess Citlalicue. Depicted as an elongated, star-studded being, Citlalicue was known as the Goddess of the Stars and the mother of several gods, among them the major God Quetzalcoatl, who was associated with the planet Venus (Milbrath, 1988; 2013: 80n40). In Figures 7 and 8 I show two more ways in which the Milky Way is depicted by other cultures who, like the Mesoamerican Tlaxacaltec people, have no connection to ancient Egypt.

Figure 7 shows a Navajo sandpainting ('dry painting' is a better term, as these paintings are made of pollen as well as various types of sand) of Mother Earth (*Sa'a Nagháí*) and Father Sky (*Bik'e Hózhó*). To contain the power of these beings, they are hemmed in by rainbows on either side. At the center of Mother Earth's body is a lake from which radiate the roots of the four sacred plants: corn, tobacco, beans, and squash. Father Sky's body is painted with the orbs of the Sun (yellow) and Moon (grey), both of which are adorned with horns of power. Significant constellations cover Father Sky's body. These include the Milky Way (*Yikáísdáhá*, lit. "that which awaits the dawn"; Maryboy and Begay, 2010: 55), which is drawn as an undulating zigzag pattern across Father Sky's arms and shoulders. Similar depictions of the Milky Way as a zigzag pattern (drawn straight, arched, or undulating) are found in sand paintings of the night sky (e.g., Newcomb and Reichard, 1937: Plate 18) or sand paintings that feature Black God (*Háshjéshjin*), a supernatural being who plays a part in placing the constellations in the sky (e.g., Haile, 1947: Plate 8). The similarities between Nut's cosmological vignette and the sand painting of Mother Earth and Father Sky are immediately apparent; simply substitute Nut for Father Sky and Geb for Mother Earth. In a similar vein, substitute the undulating curve drawn across Nut with the undulating zigzag pattern drawn across Father Sky's arms and shoulders.

The Milky Way is also depicted as an undulating zigzag pattern on the face of *Soongwuqa*, a Hopi *katsina tihu* (commonly referred to as a '*kachina* doll') that represents the Milky

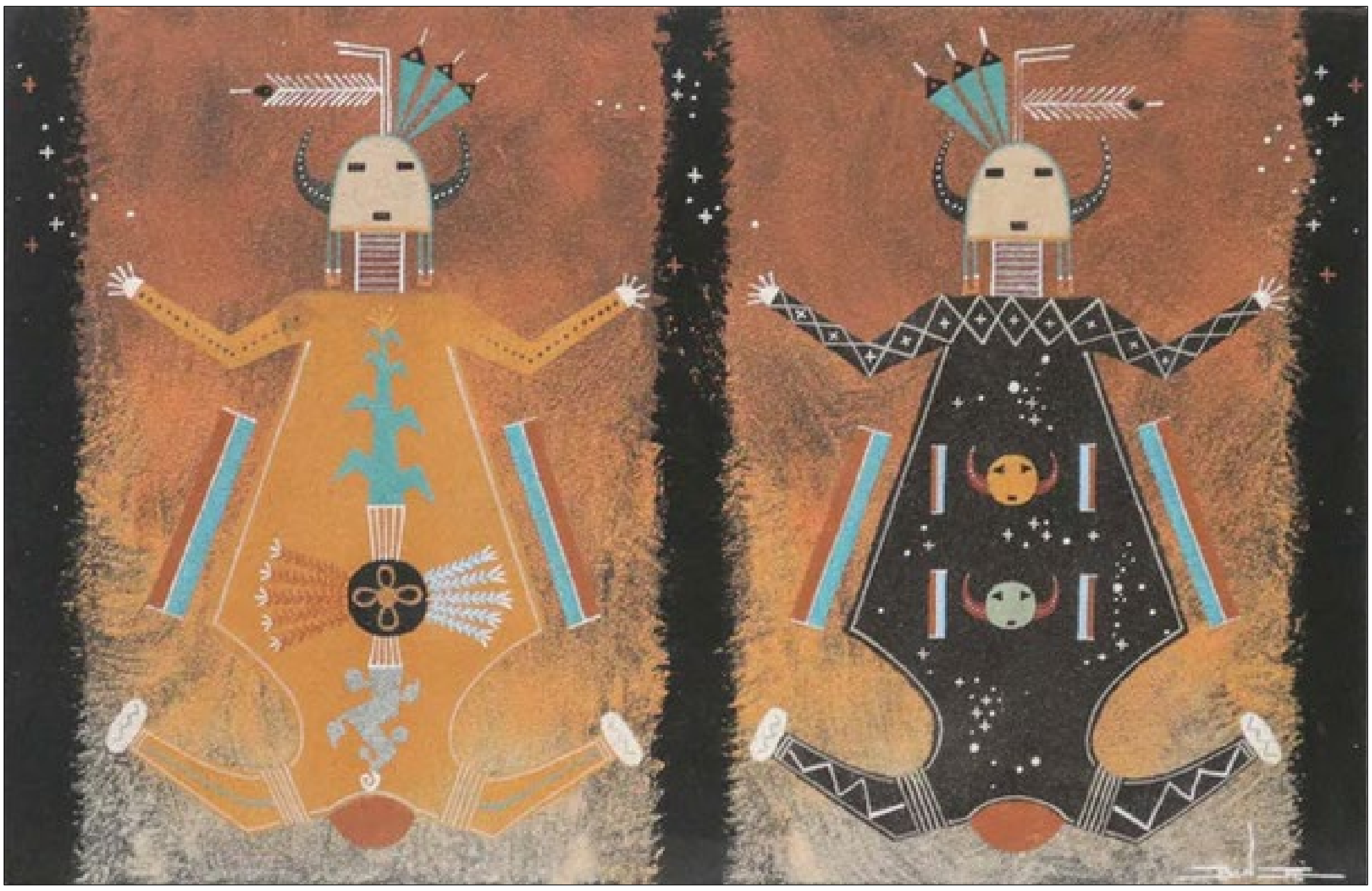

Figure 7: Navajo sand painting of Mother Earth (left) and Father Sky (right). The Milky Way is the zigzag pattern running along Father Sky's arms and shoulders (courtesy: Joe Ben, Jr; from the author's collection).

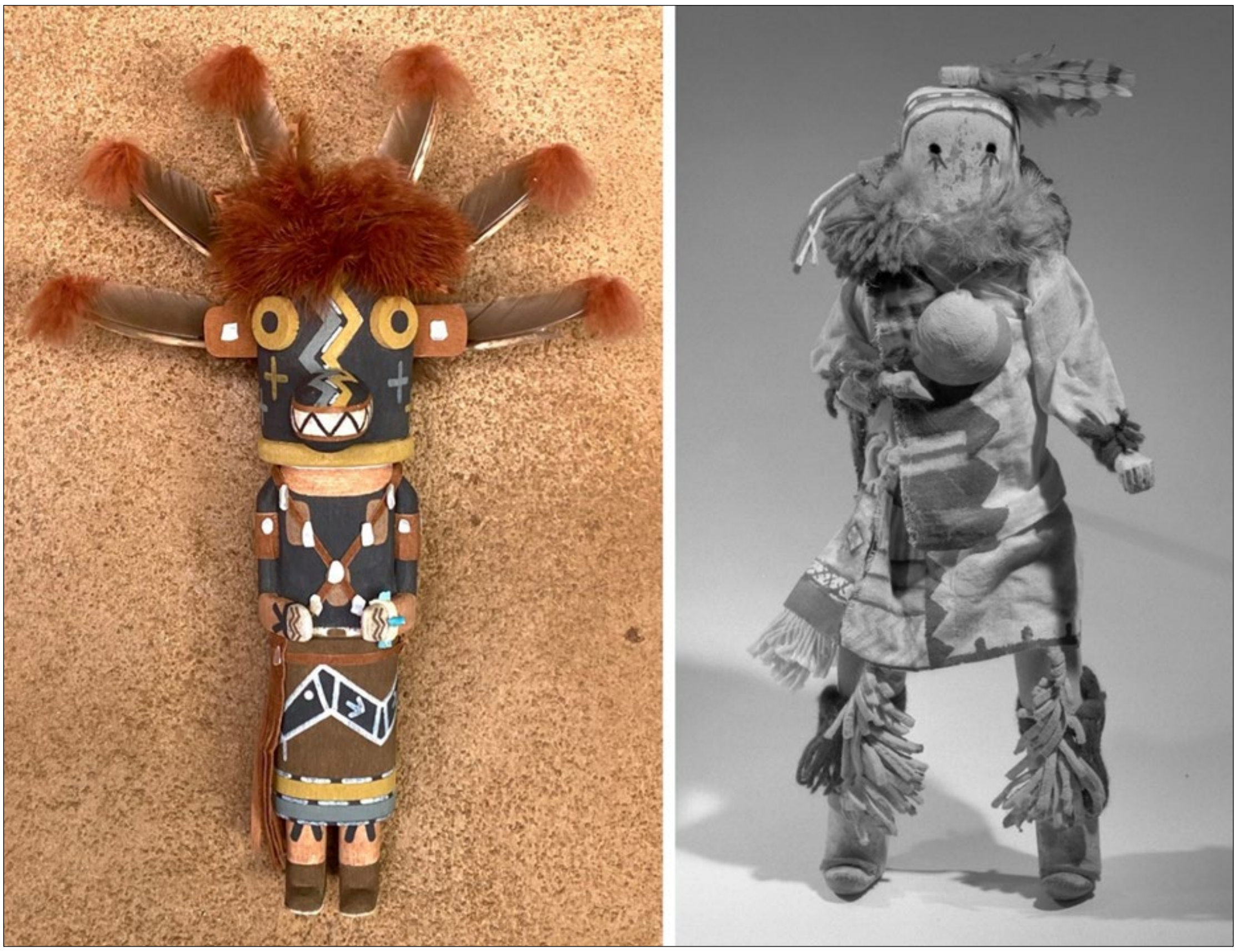

Figure 8 (Left): *Soongwuqa*, a Hopi *katsina* of the Milky Way. The Galaxy is represented by the blue and yellow zigzag pattern running down *Soongwuqa's* face (courtesy: Augustine Mowa III; from the author's collection). (Right): *Kiaklo*, a Zuni *katsina*. The Milky Way is the black-and-white checkered curve on top of *Kiaklo*'s head (courtesy: Brooklyn Museum 03.325.4614; credit: Museum Expedition 1903, Museum Collection Fund).

Way *katsina* (Figure 8). For a description of the nature of the *katsinam* and their roles in Hopi ceremonials, see Secakuku (1995: 3–4) and Wright (1977: 1–26). The exact nature of *Soongwuqa* and its role in Hopi cosmology is a subject of ongoing research (Graur, in prep.). For the purpose of this paper, it is sufficient to note that two independent Hopi carvers (Augustine Mowa III and Sean Macias) have created dolls of this *katsina*, and that both dolls feature the same face markings. The stars and crescent painted on the cheeks of the *katsina* are seen on other *katsina* dolls, such as those of the *Holokasinam* (Secakuku, 1995: 48) or *Ho'tem* families (Secakuku, 1995: 56). Similarly, snake kilts are found on many other *katsinam*. The feature that sets *Soongwuqa* apart from other *katsinam* is the colorful zigzag pattern running down its face and across its snout. In Augustine Mowa III's version (shown here) the zigzag pattern is colored blue, dashed white over a black background (which might represent the Great Rift), and yellow, while in Sean Macias's version (not shown here but available by request), the zigzags are colored red, green, white, yellow, and red. It is this zigzag pattern that defines *Soongwuqa* as the spirit of the Milky Way.

Figure 8 also shows an example of *Kiaklo*, a Zuni *katsina* (described by Bunzel, 1932: 980–985, Plate 28). Here, the Milky Way is depicted as the black-and-white checked arc across the top of the *katsina*'s head. Similar checked arcs representing the Milky Way were also documented on the faces of other Zuni *katsinam*: *Łe'lacoktipona* (Bunzel, 1932: 990–991, Plate 28) and *Käna˙kwe Mosona* (Bunzel, 1932: 1009–1011, Plate 33).

The Navajo, Hopi, and Zuni reside in modern-day Arizona and New Mexico. While their cosmogonies, origin stories, and ceremonials are not identical, cultural diffusion has resulted in certain shared elements (such as certain origin stories and the reverence of *katsinam*; Graur, in prep.). Hence, it is not surprising to see similar patterns representing the Milky Way on the bodies of important spiritual beings. However, there is no known connection between these Native American nations and the ancient Egyptians. Hence, while the similarities between the Milky Way patterns on Father Sky, *Soongwuqa*, and *Kiaklo* and the undulating curve across Nut's body may be coincidental, they may also stem from a shared human imagination, the same imagination that has caused cultures across the world and throughout history to time and again conceptualize the Milky Way in the same terms: as a river, a road, a deity, an animal, or a conduit between this world and the next (Graur, 2024b).

## 5 TRENDS IN NUT'S DEPICTIONS ON ANCIENT EGYPTIAN COFFINS

While searching for the cosmological vignette and other depictions of Nut on Egyptian coffins, I noticed two interesting trends that, along with the Milky Way feature discussed in the previous sections, shed new light on the relationship of the ancient Egyptians with the sky and on the interplay between Nut's cosmological and eschatological roles in the context of the coffin as a conduit to the afterlife.

### 5.1 A Preference for the Day Sky in the Cosmological Vignette

Among the 118 cosmological vignettes in Table 2, Nut is covered in stars in only 28 elements (24%). If we assume that a naked Nut represents the day sky and that a star-studded Nut represents the night sky, then we would expect Nut to be covered in stars 50% of the time. To ascertain whether the number of star-studded depictions of Nut in Table 2 is consistent with this hypothesis, we can treat Nut's depiction as we would a coin toss. The probability of a coin toss resulting in either heads or tails is 50%, but if we were to toss a coin 118 times, the number of heads would not always equal the number of tails. In fact, if we conducted the same experiment—flipping a coin 118 times—thousands of times, in 99.7% of those experiments, the coin would land on heads 43 to 74 times. If the coin landed on heads fewer than 43 times or more than 74 times in a given set of 118 coin tosses, we would say that the null hypothesis—that this is a fair coin—could be rejected at a statistical significance of $3\sigma$. In most of the sciences, this statistical significance is the threshold for rejecting an hypothesis or declaring the detection of a new phenomenon. Seeing Nut covered in stars in only 28 of 118 vignettes is analogous to the coin coming up heads 28 times or fewer in 118 tosses. That is expected to happen in only $4.5 \times 10^{-7}$% of experiments (equivalent to a *p*-value of $4.5 \times 10^{-9}$). This highly unlikely scenario allows us to reject the null hypothesis that the coin is fair—that the probability of finding Nut depicted with or without stars is equal—at a statistical significance of $>5\sigma$ (for comparison, the Higgs boson was declared to have been discovered once it was detected with a statistical significance of $>5\sigma$).

The same trend is seen in the cosmological vignettes in Table 3, in which Nut is covered in stars in only one of five examples (C121; Figure 9). However, although this is consistent with the fraction of star-studded vignettes in Table 2, the small size of the second sample renders it statistically insignificant.

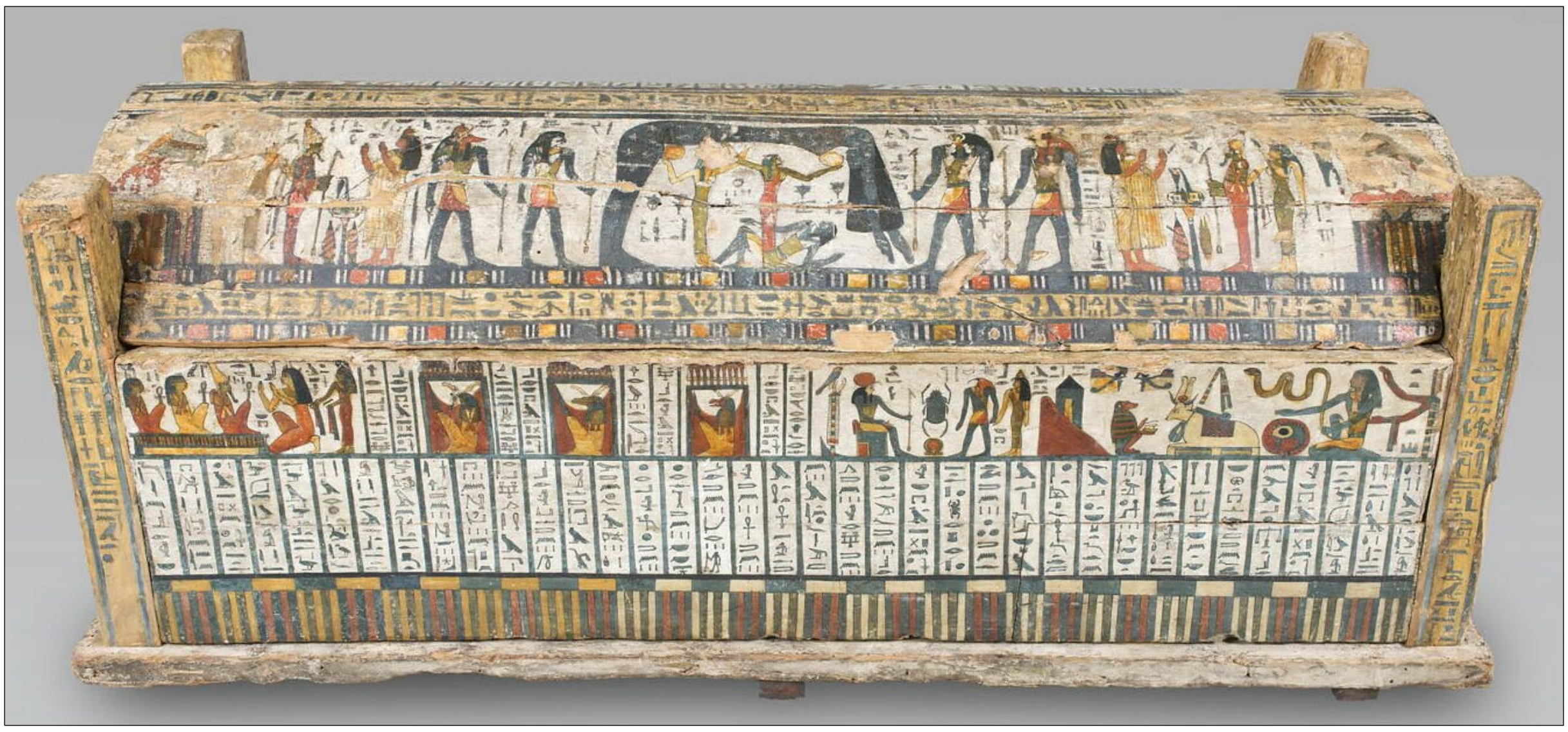

Figure 9: In the cosmological vignette on the lid of the 25th Dynasty rectangular coffin of Tachepenkhonsou (C121), Nut is colored blue and studded with stars (courtesy: Louvre E3913; © 2020 Musée du Louvre, Dist. GrandPalaisRmn / Christian Décamps).

An important caveat is that even though the sample of cosmological vignettes assembled in Table 2 is complete in the sense that it matches the catalogs assembled by Andrzej Niwiński and Agnes Klische, it is not complete in the sense that it is not a subsample of all yellow coffins. An unknown fraction of 21st/22nd Dynasty coffins has been lost or sold over the years (Niwiński, 2021), and it is possible that some of them were decorated with the cosmological vignette. There may also be more 21st/22nd Dynasty coffins buried in unknown tombs. Furthermore, as noted in Section 2.1, it is unclear whether or not Nut is covered in stars in certain vignettes. However, the size of the current sample is large enough that, given the null hypothesis, Nut would have had to be covered in stars in at least 43 vignettes (see above). One might quibble whether certain vignettes in Table 2 include stars or not, but there are only a handful of such examples, far fewer than the 15 needed to bridge the gap between 28 and 43 vignettes.

I find no discernible pattern among the coffins that are adorned with a star-studded Nut or among those in which Nut is shown naked. Some coffin sets bear multiple depictions of the cosmological vignette, but again, there is no pattern to how Nut is depicted on the inner and outer coffins of those sets. In some cases, Nut is naked on both coffin elements (C40+C41, C60+C61, C101+C102+C103), in some she is naked in one element yet covered with stars in the other (e.g., C9+C10, C29+C30, C31+C32, C89+C107), while in others she is decorated with stars on both the inner and outer coffins (e.g., C5+C6, C16+C17, C27+C35, C44+C45, C84+C85).

I suggest two possible explanations for the scarcity of vignettes in which Nut is covered in stars: either the stars are but a flourish with no deeper meaning, or the ancient Egyptians were exhibiting a marked preference for the day sky. It will be interesting to see whether a thorough study of the full-length depictions of Nut inside coffins from the 22nd Dynasty down to the Roman era (see below) will uncover a similar fraction of figures covered in stars.

## 5.2 The Cosmological Vignette and Nut's Full-Body Portrait

Though the coffin sample in Table 1 is far from complete (see Section 2), it is large enough to provide some insight into the evolution of Nut's depictions on coffins throughout Egyptian history. The earliest coffins in this sample originate in the Old Kingdom (4th Dynasty), First Intermediate Period (9th–11th Dynasties), and the Middle Kingdom (12th Dynasty). These coffins feature no visual depiction of Nut, although they do include a written invocation for the Sky Goddess to protect the deceased. Next come coffins from the 18th Dynasty, which marks the beginning of the New Kingdom. Though several deities are depicted on the sides of these anthropoid coffins, Nut is not among them. However, the lids of several of these coffins are adorned with a vulture with outspread wings. This is probably a depiction of the Goddess Nekhbet, but could also be a depiction of Nut since, along with her anthropomorphic representation, she is also represented at times as a cow, a sow, a hippopotamus, or a vulture (Billing, 2002: 13–24). Instead of a vulture, the coffin of Merymose (British Museum EA1001) sports a figure of a kneeling woman with outspread wings, who is identified as Nut. A similar description appears in the record of the sarcophagus of Djehutymose (British Museum EA1642), but a photograph of this figure is not available. Though not in my sample, coffin TT 1370, described by Bruyère (1937: Figure 10) as originating in the reign of Queen Hatshepsut of the 18th Dynasty, also bears a kneeling, winged depiction of Nut.

Clear visual depictions of Nut as a kneeling, winged goddess begin to appear in my sample on coffins from the 19th Dynasty. On both the outer and inner coffins of Khonsu (Metropolitan Museum of Art 86.1.1a, b and 86.1.2a, b) she is directly identified by her name, 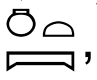, which appears next to her figure. The same named figure appears on the sarcophagus lid of Setau (British Museum EA78), and a similar, though nameless, figure appears on the lid of the coffin of Weretwashet (Brooklyn Museum 37.47Ea–d). In my sample, this popular depiction of Nut continues to appear on coffins from the 21st, 25th, and 26th Dynasties, as well as from the Ptolemaic and Roman Periods. On some coffins, the kneeling, winged depiction of Nut also appears on the exterior of the headboard (e.g., C102). Amulets with this depiction have been found either sewn into or resting on mummy wrappings (e.g., British Museum EA66940, Metropolitan Museum of Art 26.7.982a–c). This depiction is understood to be a visual invocation of spells in which Nut is called upon to spread herself in protection over the deceased, such as spells 427, 588, and 593 from the *Pyramid Texts*, or spell 181 from the *Book of the Dead* (Faulkner, 1972: 180): "… Your mother Nut has put her arms about you that she may protect you, and she will continually guard you …"

Depictions of Nut as a full-length standing figure begin to appear on the interiors of coffins as early as the 18th Dynasty: a carved depiction of a clothed Nut with her arms raised to shoulder height appears inside the lid of the coffin of Rai (Egyptian Museum in Cairo, CG 61004; Cooney, 2024: 141) and on the exterior upper quarter of the lid of a sarcophagus originally made for Queen Hatshepsut but recut for her father, King Thutmose I (Brooklyn MFA 04.278.1–2). Other examples include the 18th/19th Dynasty coffins of Setnakht and Ramesses III (Egyptian Museum in Cairo, CG 61039 and 61040, respectively; Cooney, 2024: 271, 273) and the 18th Dynasty coffin MM 13940 in the Medelhavsmuseet (Dodson, 2015: 10–11).

Two interesting depictions of Nut appear in sarcophagi belonging to the 19th Dynasty King Merneptah (one in Burial chamber J of tomb KV 08 and the other usurped by the 21st Dynasty King Psusennes I; Egyptian Museum in Cairo J87297). In both cases, Nut appears in relief on the interiors of the sarcophagi lids and she is portrayed face-on with arms raised above her head. In the sarcophagus usurped by Psusennes I, Nut wears a long dress covered in stars.

Full-length depictions of Nut continued to appear inside coffins from the 21st, 22nd, 25th, 26th, and 30th Dynasties, as well as the Ptolemaic and Roman periods (Figure 10). Most of the time, Nut appears fully clothed (in a simple dress without stars), her head turned sideways, and her arms either resting by her sides or held up to shoulder height (as on the sarcophagus made for Thutmose I). In some coffins, Nut appears twice, once with her arms lowered and once with arms raised to shoulder height (e.g., Museo Egizio S. 5228). During the 21st Dynasty, she is also winged, as in her kneeling form; these wings disappear in later dynasties. I regard the winged depictions of Nut inside 21st Dynasty coffins as a further invocation of Nut's protection of the deceased.

The naked, star-studded version of Nut's full-body portrait is exceedingly rare during the 21st and early 22nd Dynasties. I know of only one example from the 21st Dynasty, on the floorboard of the outer coffin of Tanetschedmut (Louvre N2594; Niwiński, 2023: 470; Niwiński and Rigault-Déon, 2024: 546). Although my sample of late 22nd Dynasty coffins is small, in all 11 examples, Nut is painted identically: dressed in either a long skirt or dress; usually with heavy, bare breasts; and facing forward with her arms raised to shoulder height. In the 25th Dynasty coffins in my sample, Nut is clothed in a full-length dress, but her breasts are still exposed. As in the 22nd Dynasty portraits, she faces forward, but now her arms are raised above her head to grasp the Sun. From the 26th Dynasty onwards, Nut's portrait resembles the one seen on the sarcophagus usurped by Psusennes I from Merneptah: she is fully naked, front-facing, and her arms are raised above her head. In many of these portraits, one or more Solar discs are painted along her body, and she is either covered or surrounded by stars (during the Ptolemaic and Roman periods, the stars are accompanied by visual representations of Egyptian constellations and the Greek zodiac).

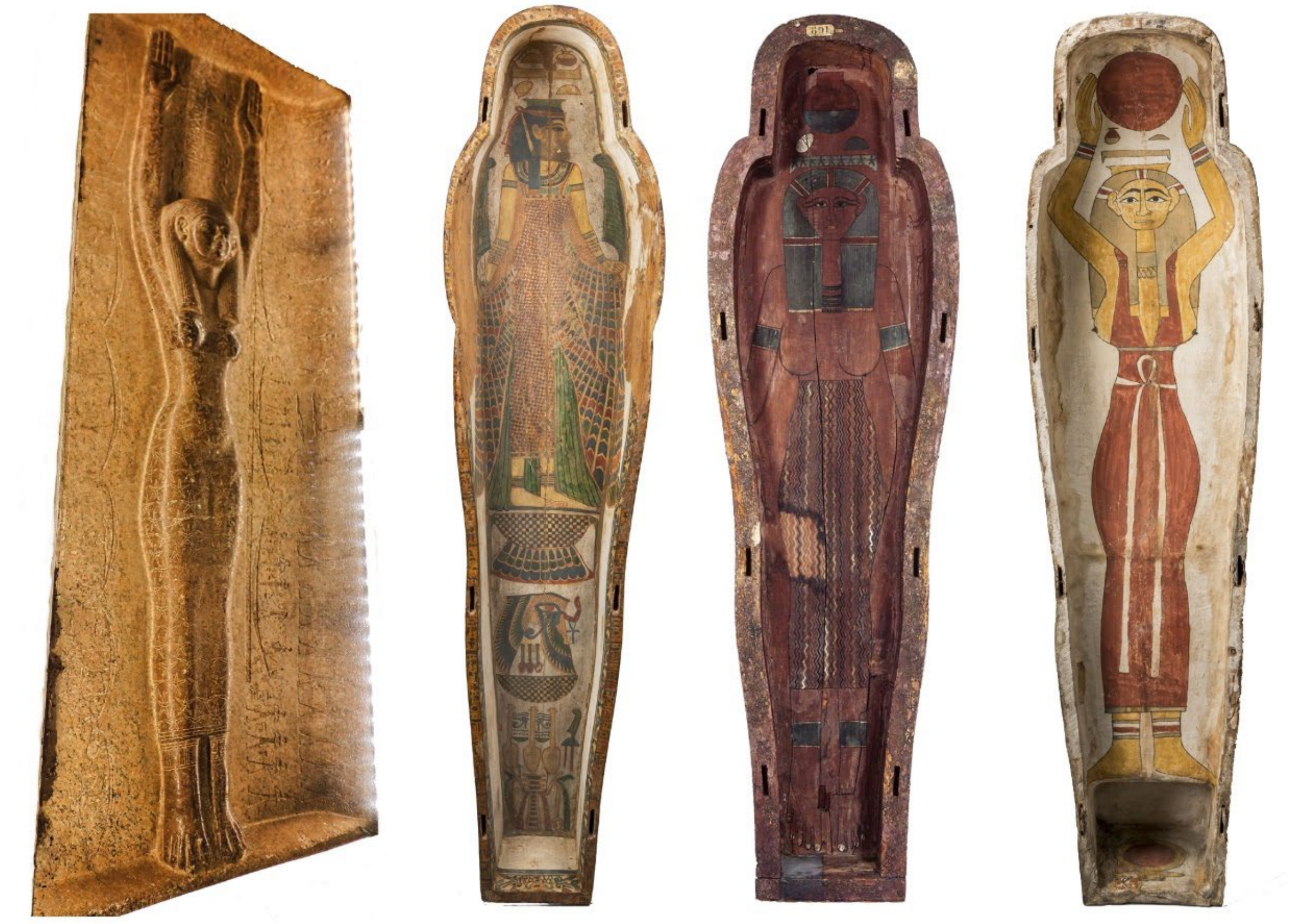

Figure 10a: The evolution of Nut's full-length portrait. From left to right: 19th Dynasty red granite sarcophagus of King Merneptah, usurped by 21st Dynasty King Psusennes I (courtesy: Egyptian Museum in Cairo J87297; image credit: CK-TravelPhotos/Shutterstock.com); 21st Dynasty inner coffin of Butehamon (courtesy: Museo Egizio 2237); 22nd Dynasty coffin of Tanetmit (courtesy: Louvre N2587; © 2004 Musée du Louvre / Georges Poncet); 25th Dynasty coffin of Tariri (courtesy: Museo Egizio 2220).

After the 21st Dynasty, the vignette in which Nut appears as the Goddess of the Sycamore also evolves into a third version of her full-length portrait, in which she is painted with a fruit-bearing tree in the background. In some of these portraits, she is seen pouring water for *ba* birds on either side of her (e.g., Fitzwilliam Museum E.2.1869).

The evolution of Nut's full-length portrait (Figure 10) may explain the popularity of the cosmological vignette on 21st/22nd Dynasty coffins and its subsequent disappearance in later dynasties. I suggest that Nut's full-length portraits served two different purposes. The clothed version (including the sycamore variant) evoked Nut's role as a guardian of the deceased. This is stressed in 18th and 21st Dynasties versions where Nut is winged. On the other hand, Nut's naked, cosmological variant is a variation of Nut's depiction in the Ramesside monumental versions of the *Fundamentals* ... It may even be a visual evocation of the even earlier *Tale of Sinuhe*, in which the King promises Sinuhe "… a mummy case of gold, a mask of lapis lazuli, a heaven over you …" (Parkinson, 1997: 36). Although Shu and Geb are absent from Nut's full-length portraits, her outstretched arms and legs evoke her arched form, painted to express the arched sky as it would be viewed by the recumbent figure of the deceased looking up at the sky. This is further stressed in images where Nut's hair seems to stand on end; it is really flowing down from her bent head (as also seen in several of the cosmological vignettes, e.g., Figure 1d). The appearance of the Sun, stars, various constellations, or the zodiac stresses the cosmological nature of these depictions.

During the 21st Dynasty, the cosmological vignette made Nut's cosmological full-length portrait redundant. This is seen numerically in my sample, where Nut's cosmological vignette appears on 118 (41%) of the 287 yellow coffins but her full-length portrait appears on only 17 (6%) of the coffins. Nine (53%) of the latter coffins display both the cosmological vignette and a full-length guardian portrait, perhaps revealing a special affinity for Nut among the makers or users of those coffins. Note that the fraction of cosmological vignettes among yellow coffins presented above is an overestimate, as the sample of coffins in Table 1 was assembled as part of an explicit search for the cosmological vignette. However, the relative fraction of full-length portraits relative to cosmological vignettes is still of interest.

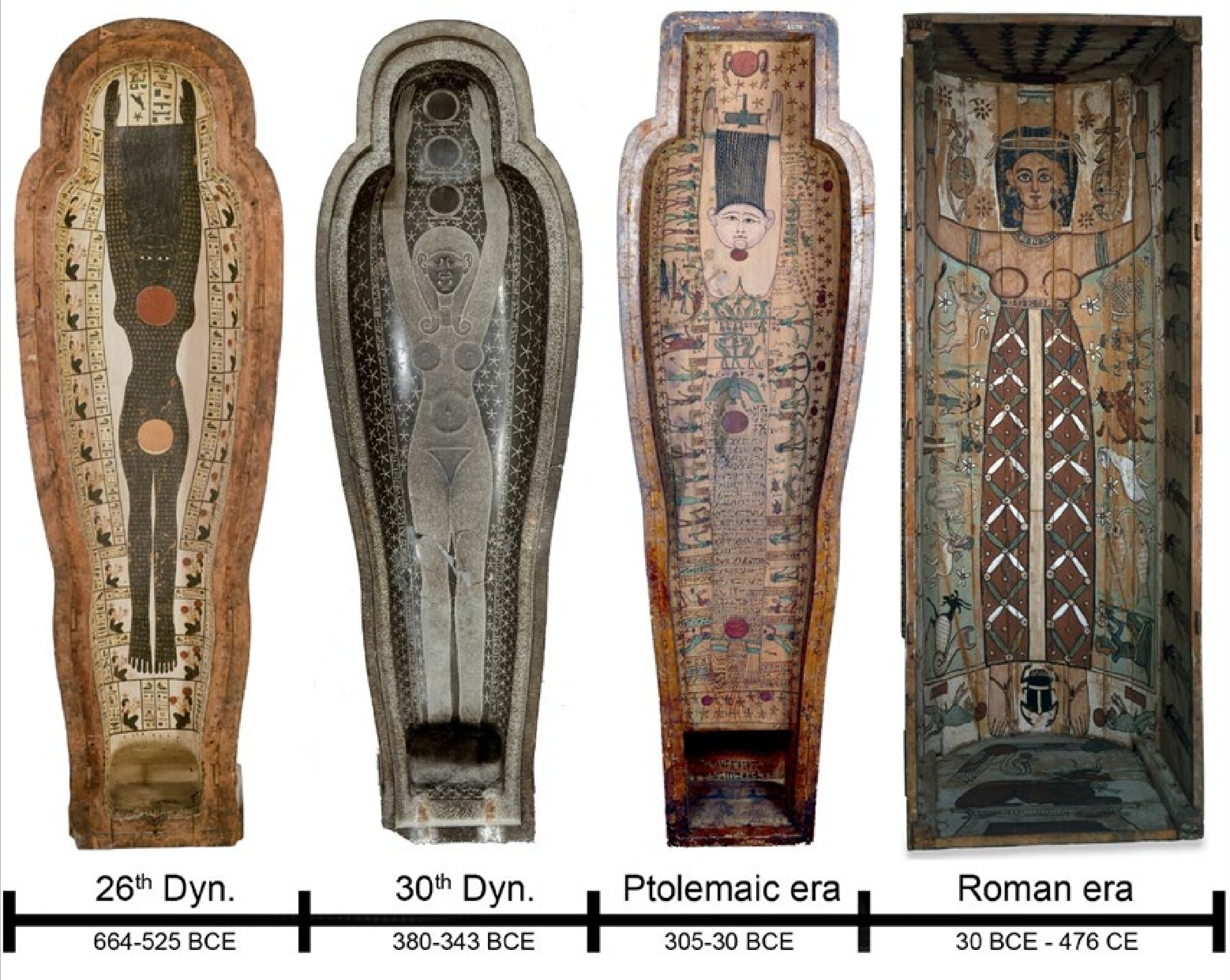

Figure 10b: The evolution of Nut's full-length portrait (continued). From left to right: 26th Dynasty coffin of Peftjaoeneith (courtesy: Rijksmuseum van Oudheden AMM5-e); 30th Dynasty sarcophagus of Tenthapi (courtesy: Louvre E84; © 2006 Musée du Louvre, Dist. GrandPalaisRmn / Georges Poncet); Ptolemaic-era coffin of Hornedjitef (courtesy: British Museum EA6678; © The Trustees of the British Museum. Shared under a Creative Commons Attribution-NonCommercial-ShareAlike 4.0 International (CC BY-NC-SA 4.0) licence); Roman-era coffin of Soter (courtesy: British Museum EA6705; © The Trustees of the British Museum. Shared under a Creative Commons Attribution-NonCommercial-ShareAlike 4.0 International (CC BY-NC-SA 4.0) licence).

For whatever reason, later dynasties preferred Nut's cosmological full-length portrait over the cosmological vignette. Thus, while the cosmological vignette with Nut, Shu, and Geb all but disappeared after its heyday on the sides of yellow coffins, Nut's cosmological role in the transition of the deceased to the afterlife remained prominent. The fact that some coffins displayed both guardian and cosmological portraits of Nut (e.g., Metropolitan Museum of Art O.C.22a, b; Medelhavsmuseet NME 007) or sycamore and cosmological portraits (e.g., British Museum EA6705 and EA6706) shows that Nut's multiple eschatological and cosmological roles remained crucial to the functions of the coffin as a vehicle for the deceased's transition to the afterlife. However, this observation needs to be substantiated with a larger, unbiased sample of coffins.

## 5.3 The Orientation of Nut's Arms

A final note should be made about the orientation of Nut's arms in her cosmological full-length portraits. As noted above, her arms are always shown stretched straight over her head. This orientation clashes with two suggestions in the literature for how Nut's body, which is usually only studied in the two-dimensional de-pictions on the cosmological vignettes, should be reconstructed in three dimensions.

One suggestion is that Nut's arms and legs constituted the four cardinal points (e.g., Niwiński, 2011). However, this spread-eagled image of Nut clashes both with her full-length portraits and with her textual description in the *Fundamentals* …, which describes Nut as (von Lieven, 2007): "Her rear is in the east, and her head is in the west" and "Her right arm is on the northwest side, [her left arm] is on the southeast side." While this description rules out the first suggestion, it forms the basis for the second suggestion, made by Graur (2024a), that the description of Nut's arms was seen as evidence for connecting Nut to the Milky Way, as the latter assumes the same Northwest–Southeast orientation in the Egyptian night sky during the winter.

One cannot argue that the floorboard of the coffin did not provide enough space to depict Nut's arms with the orientation described in the *Fundamentals* … While most of Nut's full-length portraits are restricted to a single plane, there are ample examples where Nut's arms occupy both the floorboard and the inner sides of the coffins. On 22nd Dynasty coffins, for example, Nut's arms seem to terminate at the elbows, but the rest of her arms, raised to shoulder height, are painted on either side of the floorboard. Instead, it is possible that, just as Nut's full-length portrait evolved over time, so did the Egyptians' mental image of Nut. Our oldest description—visual and textual—of Nut is found in the witness of the *Fundamentals …* in the 18th-Dynasty Cenotaph of Seti I. However, von Lieven (2011) argues that the ephemerides below Nut's body date to the 12th Dynasty and that the list of stars along her back may have originated in the earlier Old Kingdom. Thus, it is plausible that when the *Fundamentals* … was originally composed, Nut's arms were imagined to orient along the arch of the Milky Way. But, by the time the *Fundamentals* … was inscribed in the Cenotaph of Seti I, her visual image had evolved away from the original (or at least preceding) image. This new two-dimensional image was then given three-dimensional form in the full-length coffin portraits described above.

## 6 CONCLUDING REMARKS

By collecting and analyzing examples of Nut's cosmological vignettes on yellow coffins from the 21st and early 22nd Dynasties, I have come up with three new insights. First, I have highlighted a unique feature of the cosmological vignette on the outer coffin of Nesitaudjatakhet. On this coffin, Nut is decorated not only with stars but also with a thick, black undulating curve that bisects her entire body. This curve resembles the Great Rift of the Milky Way and is similar to depictions of the Milky Way as undulating zigzag or checkered patterns on the bodies of Navajo, Hopi, and Zuni spiritual beings. It is also similar to undulating curves seen bisecting the astronomical ceiling in the tomb of Seti I and along Nut's back in the witnesses of the *Books of the Day and the Night* in the tombs of Ramesses IV, VI, and IX. Based on these similarities, I suggest that this undulating curve is the visual representation of the Milky Way, which the Egyptians saw as cleaving the sky in two. Moreover, it reopens the possibility of 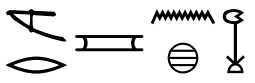 (*mr-nḫ3* – Winding Waterway) being the ancient Egyptian name of our Galaxy.

The undulating curve provides a further link between Nut and the Milky Way, on top of the textual link presented by Graur (2024a). However, the fact that this feature is observed in only one coffin further strengthens the suggestion first made by Graur (2024a) that, although the Milky Way may have been linked to Nut, the two were not synonymous with each other. We should not think of the Milky Way as a representation of Nut but as one more astronomical phenomenon that, like the Sun and the stars, was part of the sky and thus could appear on Nut's body.

Second, only a quarter of the cosmological vignettes depict Nut as covered in stars. If we interpret Nut's naked form as the day sky and her star-studded form as the night sky, the under-representation of the night sky may indicate an inherent preference for the day sky. A thorough examination of funerary papyri and of Nut's full-length portraits inside coffins is required to test whether the under-representation of starry depictions of Nut continued to manifest itself throughout ancient Egyptian history.

Third, while the cosmological vignette was already understood to be a continuation of Nut's depiction in the monumental versions of the *Fundamentals* … found in tombs from preceding dynasties, I argue that it continued to evolve after the 21st/22nd Dynasties, when it merged with the full-length portraits found inside many coffins all the way to the Roman era. Nut's cosmological portrait, which may have originated as early as the 19th Dynasty, was briefly overshadowed in the 21st and early 22nd Dynasties by the cosmological vignette. Later dynasties saw the disappearance of the cosmological vignette and the reappearance of the cosmological full-length portrait. The mix-ture of guardian, sycamore, and cosmological versions of Nut's full-length portrait stresses the importance of her combined cosmological and eschatological roles to the nature of the coffin as a conduit to the afterlife. The orientation of Nut's arms in her cosmological portraits rules out the suggestions that her arms and legs rested on the four cardinal points and challenges the textual Northwest–Southeast orientation of her arms in the *Fundamentals* ... The latter may point to an evolving conception of Nut's three-dimensional form and her link to the Milky Way.

Finally, the catalogs assembled here underline the importance of fully digitizing museum catalogs and providing free access to them through public-facing websites. I am deeply grateful to those museums that have already made their collections accessible in this manner and I urge other museums (and the governments and private foundations that fund them) to create similar digital collections. The fire that gutted the Museu Nacional in Rio de Janeiro in 2018 is a reminder that these efforts should be carried out sooner rather than later.

## 7 ACKNOWLEDGMENTS

I thank Mykola Tarasenko, Agnes Klische, Andrzej Niwiński, Sarah Symons, and the anonymous referee for fruitful discussions and comments, as well as for sharing their data and catalogs with me. I also thank Osama Fathi for sharing his photographs of the Milky Way above Egypt. Several Egyptologists and curators have provided me with images and helpful information about the coffins under their care: Kerry Gnandt (The Cleveland Museum of Art), Lisa Graves (Bristol Museum & Art Gallery), Lisa Haney (Carnegie Museum of Natural History), Michaela Huettner (Kunsthistorisches Museum Wien), Marcin Karczewski (National Museum of Warsaw), Alexandra Küffer (Swiss Coffin Project), Vincent Lecourt (Musée Dobrée), Noémie Monbaron (Musée d'Art et d'Histoire in Geneva), Annie Shanley (Michael C. Carlos Museum), Susanne Töpfer (Museo Egizio / U. Pisa), and Ludmila Werkström (Museum Gustavianum). Finally, I thank The Trustees of the British Museum for permission to use their images of the coffins of Hornedjitef (EA6678) and Soter (EA6705). These images are shared under a Creative Commons Attribution-NonCommercial-ShareAlike 4.0 International (CC BY-NC-SA 4.0) licence.

This research has made extensive use of the libraries of the University of Cambridge, University College London, the British Museum, and the American Museum of Natural History, as well as the Internet Archive. It has also made use of the JSesh open-source hieroglyphics editor (Rosmorduc, 2014). Supporting research data are available on reasonable request from the corresponding author.

**Dr. Or Graur** is an Associate Professor of Astrophysics at the University of Portsmouth's Institute of Cosmology and Gravitation. He is also an Honorary Associate Professor at University College London and a Research Associate at the American Museum of Natural History. Dr. Graur completed his PhD in Physics and Astronomy at Tel Aviv University in 2013. Following his graduate studies, he worked with Professor Adam Riess at Johns Hopkins University (2013–2014) and Professor Maryam Modjaz at New York University (2014–2016). He then won an independent NSF Astronomy and Astrophysics Postdoctoral Fellowship, which he took to the Center for Astrophysics | Harvard & Smithsonian (2016–2020).

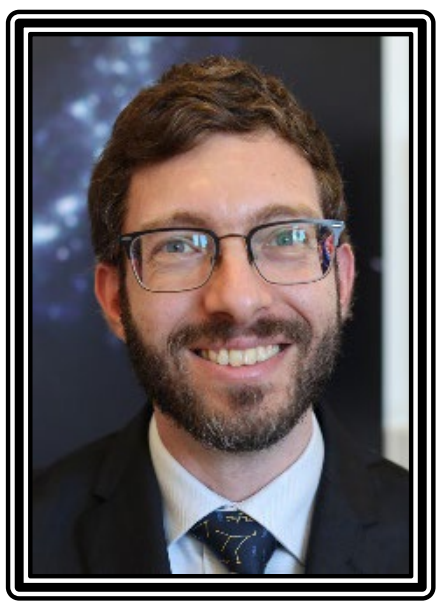

Or conducts observational studies of supernovae and tidal disruption events. His main interest is figuring out the progenitors of Type Ia supernovae, which he pursues by measuring supernova rates, studying correlations with host-galaxy properties, and using the Hubble Space Telescope to observe nearby supernovae when they are hundreds of days old. His group also studies the host galaxies of tidal disruption events and the connection between these flares and rare extreme coronal-line emitting galaxies. He is also passionate about education and public outreach: in 2016, he founded the Harvard Science Research Mentoring Program, which sees early-career astronomers supervise high-school students in year-long independent research projects. The program, which he directed between 2016 and 2019, is still in operation.

Since 2011, Or has authored >80 peer-reviewed papers on these subjects, as well as two popular-science books: *Supernova* (2022, MIT Press) and *Galaxies* (2024, MIT Press). These books kindled an interest in the history of astronomy and in cultural astronomy. The current paper is the second product of an ongoing project to study the multicultural mythology of the Milky Way.